\documentclass[journal]{IEEEtran}

\usepackage[T1]{fontenc}
\usepackage{amsmath,amssymb,amsfonts,mathrsfs,bm}
\usepackage{amsthm}
\usepackage{cite}
\usepackage{array}

\usepackage[shortlabels]{enumitem}
\usepackage{graphicx}
\usepackage{url}
\usepackage{color}
\usepackage{algorithm,algorithmic}
\usepackage{multirow}
\usepackage[table]{xcolor}
\usepackage[normalem]{ulem}
\usepackage[style=base]{caption}
\usepackage{array}
\newcolumntype{L}[1]{>{\raggedright\let\newline\\\arraybackslash\hspace{0pt}}m{#1}}
\newcolumntype{C}[1]{>{\centering\let\newline\\\arraybackslash\hspace{0pt}}m{#1}}
\newcolumntype{R}[1]{>{\raggedleft\let\newline\\\arraybackslash\hspace{0pt}}m{#1}}
\usepackage{xparse}

\ifx\notloadhyperref\undefined
	\ifx\loadbibentry\undefined
		\usepackage[hidelinks]{hyperref}
	\else
		\usepackage{bibentry}
		\makeatletter\let\saved@bibitem\@bibitem\makeatother
		\usepackage[hidelinks]{hyperref}
		\makeatletter\let\@bibitem\saved@bibitem\makeatother
	\fi
\else
\relax
\fi

\usepackage[capitalize]{cleveref}
\crefname{equation}{}{}
\Crefname{equation}{}{}
\crefname{claim}{claim}{claims}
\crefname{step}{step}{steps}
\crefname{line}{line}{lines}
\crefname{dmath}{}{}
\crefname{dseries}{}{}
\crefname{dgroup}{}{}
\crefname{Theorem}{Theorem}{Theorems}
\crefname{Corollary}{Corollary}{Corollaries}
\crefname{Proposition}{Proposition}{Propositions}
\crefname{Lemma}{Lemma}{Lemmas}
\crefname{Definition}{Definition}{Definitions}
\crefname{Example}{Example}{Examples}
\crefname{Assumption}{Assumption}{Assumptions}
\crefname{Remark}{Remark}{Remarks}
\crefname{Theorem_A}{Theorem}{Theorems}
\crefname{Corollary_A}{Corollary}{Corollaries}
\crefname{Proposition_A}{Proposition}{Propositions}
\crefname{Lemma_A}{Lemma}{Lemmas}
\crefname{Definition_A}{Definition}{Definitions}

\ifx\useTheoremCounter\undefined
\newtheorem{Theorem}{Theorem}
\newtheorem{Corollary}{Corollary}
\newtheorem{Proposition}{Proposition}

\else
\newtheorem{Theorem}{Theorem}

\newtheorem{Proposition}[theorem]{Proposition}
\fi

\newtheorem{Definition}{Definition}
\newtheorem{Example}{Example}

\theoremstyle{remark}

\newcommand{\calG}{\mathcal{G}}

\newcommand{\calO}{\mathcal{O}}

\newcommand{\calQ}{\mathcal{Q}}

\newcommand{\calS}{\mathcal{S}}

\newcommand{\calU}{\mathcal{U}}
\newcommand{\calV}{\mathcal{V}}

\newcommand{\calX}{\mathcal{X}}
\newcommand{\calY}{\mathcal{Y}}
\newcommand{\calZ}{\mathcal{Z}}

\newcommand{\bg}{\mathbf{g}}

\newcommand{\bx}{\mathbf{x}}

\newcommand{\by}{\mathbf{y}}

\newcommand{\bz}{\mathbf{z}}

\newcommand{\scH}{\mathscr{H}}

\DeclareSymbolFont{bsfletters}{OT1}{cmss}{bx}{n}
\DeclareSymbolFont{ssfletters}{OT1}{cmss}{m}{n}
\DeclareMathSymbol{\bsfGamma}{0}{bsfletters}{'000}
\DeclareMathSymbol{\ssfGamma}{0}{ssfletters}{'000}
\DeclareMathSymbol{\bsfDelta}{0}{bsfletters}{'001}
\DeclareMathSymbol{\ssfDelta}{0}{ssfletters}{'001}
\DeclareMathSymbol{\bsfTheta}{0}{bsfletters}{'002}
\DeclareMathSymbol{\ssfTheta}{0}{ssfletters}{'002}
\DeclareMathSymbol{\bsfLambda}{0}{bsfletters}{'003}
\DeclareMathSymbol{\ssfLambda}{0}{ssfletters}{'003}
\DeclareMathSymbol{\bsfXi}{0}{bsfletters}{'004}
\DeclareMathSymbol{\ssfXi}{0}{ssfletters}{'004}
\DeclareMathSymbol{\bsfPi}{0}{bsfletters}{'005}
\DeclareMathSymbol{\ssfPi}{0}{ssfletters}{'005}
\DeclareMathSymbol{\bsfSigma}{0}{bsfletters}{'006}
\DeclareMathSymbol{\ssfSigma}{0}{ssfletters}{'006}
\DeclareMathSymbol{\bsfUpsilon}{0}{bsfletters}{'007}
\DeclareMathSymbol{\ssfUpsilon}{0}{ssfletters}{'007}
\DeclareMathSymbol{\bsfPhi}{0}{bsfletters}{'010}
\DeclareMathSymbol{\ssfPhi}{0}{ssfletters}{'010}
\DeclareMathSymbol{\bsfPsi}{0}{bsfletters}{'011}
\DeclareMathSymbol{\ssfPsi}{0}{ssfletters}{'011}
\DeclareMathSymbol{\bsfOmega}{0}{bsfletters}{'012}
\DeclareMathSymbol{\ssfOmega}{0}{ssfletters}{'012}

\DeclareMathOperator*{\argmax}{arg\,max}
\DeclareMathOperator*{\argmin}{arg\,min}

\newcommand{\qednew}{\nobreak \ifvmode \relax \else
      \ifdim\lastskip<1.5em \hskip-\lastskip
      \hskip1.5em plus0em minus0.5em \fi \nobreak
      \vrule height0.75em width0.5em depth0.25em\fi}

\newcommand{\bzero}{\mathbf{0}}

\newcommand{\ofrac}[1]{{\frac{1}{#1}}}

\newcommand{\ip}[2]{{\left\langle{#1},{#2}\right\rangle}}
\newcommand{\norm}[1]{{\left\lVert{#1}\right\rVert}}

\newcommand{\cond}[2]{\left. {#1}\, \middle| \, {#2} \right.}

\DeclareDocumentCommand \P { g d() g } {%
	\IfNoValueTF {#3}
	{%
		\IfNoValueTF {#1}
		{%
			\IfNoValueTF {#2}
			{%
				\mathbb{P}%
			}%
			{%
				\mathbb{P}\left({#2}\right)%
			}%
		}%
		{%
			\IfNoValueTF {#2}
			{%
				\mathbb{P}_{#1}%
			}%
			{%
				\mathbb{P}_{#1}\left({#2}\right)%
			}%
		}%
	}%
	{%
		\IfNoValueTF {#1}
		{%
			\mathbb{P}\left(\cond{#2}{#3}\right)%
		}%
		{%
			\mathbb{P}_{#1}\left(\cond{#2}{#3}\right)%
		}%
	}%
}

\DeclareDocumentCommand \E { g o g } {%
	\IfNoValueTF {#3}
	{%
		\IfNoValueTF {#1}
		{%
			\IfNoValueTF {#2}
			{%
				\mathbb{E}%
			}%
			{%
				\mathbb{E}\left[{#2}\right]%
			}%
		}%
		{%
			\IfNoValueTF {#2}
			{%
				\mathbb{E}_{#1}%
			}%
			{%
				\mathbb{E}_{#1}\left[{#2}\right]%
			}%
		}%
	}%
	{%
		\IfNoValueTF {#1}
		{%
			\mathbb{E}\left[\cond{#2}{#3}\right]%
		}%
		{%
			\mathbb{E}_{#1}\left[\cond{#2}{#3}\right]%
		}%
	}%
}

\definecolor{gray90}{gray}{0.9}

\newcommand{\blue}[1]{{{\color{black} #1}}}

\newcommand{\msout}[1]{\text{\color{green} \sout{\ensuremath{#1}}}}
\newcommand{\del}[1]{{\color{green}\ifmmode \msout{#1}\else\sout{#1}\fi}}

\newcommand{\hide}[1]{}

\graphicspath{{./Figures/}}
\begingroup\expandafter\expandafter\expandafter\endgroup
\expandafter\ifx\csname pdfsuppresswarningpagegroup\endcsname\relax
\else
\fi

\usepackage{epstopdf}
\usepackage{empheq}
\DeclareGraphicsExtensions{.eps}

\usepackage[labelformat=simple]{subcaption}

\begin{document}
\title{On the Relationship Between Inference and Data Privacy in Decentralized IoT Networks}

\author{Meng~Sun, and
        Wee~Peng~Tay,~\IEEEmembership{Senior~Member,~IEEE}
\thanks{This research is supported by the Singapore Ministry of Education Academic Research Fund Tier 1 grant 2017-T1-001-059 (RG20/17).}
\thanks{The authors are with the Department of Electrical and Electronic Engineering, Nanyang Technological University, Singapore, e-mails: MSUN002@e.ntu.edu.sg, wptay@ntu.edu.sg}
}

\maketitle
\begin{abstract}
In a decentralized Internet of Things (IoT) network, a fusion center receives information from multiple sensors to infer a public hypothesis of interest. To prevent the fusion center from abusing the sensor information, each sensor sanitizes its local observation using a local privacy mapping, which is designed to achieve both inference privacy of a private hypothesis and data privacy of the sensor raw observations. Various inference and data privacy metrics have been proposed in the literature. We introduce the concept of privacy implication (with vanishing budget) to study the relationships between these privacy metrics. We propose an optimization framework in which both local differential privacy (data privacy) and information privacy (inference privacy) metrics are incorporated. In the parametric case where sensor observations' distributions are known \emph{a priori}, we propose a two-stage local privacy mapping at each sensor, and show that such an architecture is able to achieve information privacy and local differential privacy to within the predefined budgets. For the nonparametric case where sensor distributions are unknown, we adopt an empirical optimization approach. Simulation and experiment results demonstrate that our proposed approaches allow the fusion center to accurately infer the public hypothesis while protecting both inference and data privacy.
\end{abstract}


\section{Introduction}
With the proliferation of Internet of Things (IoT) devices like smart phones and home voice recognition assistants, protecting the privacy of users has attracted considerable attention in recent years \cite{roman2013features,SunTay:C16,HeTaySun:C16,he2017multi}. Data collected by IoT devices to provide services that lead to better healthcare, more efficient air conditioning, and safer cities \cite{Alemdar2010,Butun2014}, may be used for more nefarious purposes like tracking an individual without her explicit consent. An individual's privacy has been enshrined as a fundamental right through the laws of many countries \cite{gdpr,dataprotectionact2012,digital_privacy_act}, and privacy protection mechanisms are increasingly being adopted by IoT product makers. For example, Apple Inc.\ have recently started to implement local differential privacy mechanisms into their iCloud product \cite{apple_user_guide}. 

\begin{figure}[!tb]
\centering
\includegraphics[width=0.45\textwidth]{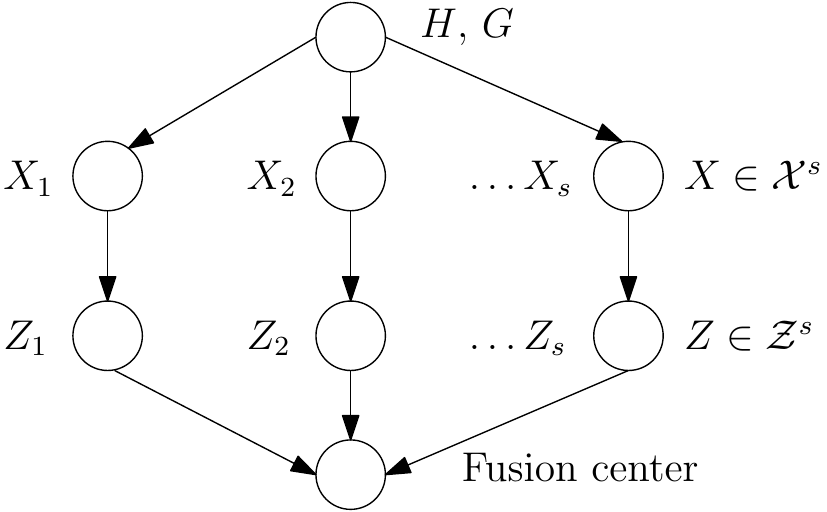}
\caption{An IoT network with public hypothesis $H$ and private hypothesis $G$. Each sensor $i$ observes a private observation $X_i$, which is first sanitized to $Z_i$ before being sent to the fusion center.}
\label{fig:PBL}
\end{figure}

We consider an IoT network (see \cref{fig:PBL}) consisting of multiple sensors, each making a private observation, which is first distorted through a privacy mapping before being sent to a fusion center. The information received from all the sensors is used by the fusion center to perform inference on a public hypothesis of interest. Privacy for this IoT network can be categorized into two classes: data privacy and inference privacy. Data privacy refers to the protection of each sensor's raw private observation from the fusion center, i.e., upon receiving information from all the sensors, it is difficult for the fusion center to infer the original sensor observations. Protecting data privacy alone is not sufficient to prevent privacy leakage. A data privacy mechanism obfuscates the raw data while still allowing statistical information to be extracted from the data. Given multiple information sources, each with its local data privacy mechanism, it is possible to perform a correlation attack \cite{liu2016dependence} leading to de-anonymization and other types of privacy leakage as shown in the examples in \cite{cormode2011personal}.

Inference privacy refers to preventing the fusion center from making certain statistical inferences it has not been authorized to perform. We call a hypothesis a public hypothesis if its inference or detection is to be achieved by the fusion center. We call a hypothesis a private hypothesis, if its true state is not authorized to be inferred by the fusion center. For example in using on-body wearables for fall detection, the fusion center is authorized to perform fall detection, but not authorized to detect if a person is exercising or performing another activity. Prevention of statistical inference of the latter activities is inference privacy, while preventing the fusion center from reconstructing the raw sensor data up to a certain fidelity is data privacy. It can be seen from this example that distortion of the raw sensor data to achieve data privacy does not necessarily remove all statistical information required to infer if the person is performing a private activity, unless the sensor data is so heavily distorted that even fall detection becomes difficult. On the other hand, inference privacy also does not guarantee data privacy as inference privacy mechanisms target to protect only specific statistical inferences. For example, blurring certain parts of an image may prevent inference of certain objects in the image, but does not necessarily distort the whole image significantly.  

The main focus of this paper is to derive insights into the relationships between various data and inference privacy metrics, and to design a privacy-preserving decentralized detection architecture for IoT networks where the level of data and inference privacy can be chosen. We aim to achieve a good tradeoff between data privacy, inference privacy and the detection accuracy of of the public hypothesis at the fusion center.

\subsection{Related Work}

\blue{Various works have focused on protecting data privacy while providing utility. In a privacy-preserving consensus network, each node share obfuscated information with each other. The papers \cite{huang2012differentially,nozari2015differentially,manitara2013privacy, braca2016learning,mo2017privacy,lazzeretti2014secure,ambrosin2017odin,hallgren2017privatepool} proposed methods that allow the nodes to obtain the correct information collaboratively without sharing their private observations. These works consider data privacy preserving methods for a fully distributed network where there is no fusion center. This is different from the IoT model that we study in this paper and is out of our current scope. Moreover, the issue of inference privacy has also not been addressed.}

In cloud services and applications, data privacy can be achieved using homomorphic encryption\cite{Gentry2009,brakerski2014leveled}, which allows a cloud server to compute on encrypted data without decryption. The encrypted result is then made available to the requester, who is able to decrypt it. By comparison, in decentralized detection, the fusion center needs to play the roles of both the cloud server and requester, making it impossible to apply homomorphic encryption techniques here. Other data privacy works propose to corrupt each sensor's local observation so that the fusion center cannot infer it\cite{wang2003using, xiong2016randomized, liao2017hypothesis}. In \cite{duchi2013local}, the authors analyzed the tradeoff between local differential privacy budget and the utility of statistical estimators used at the fusion center. The paper \cite{wang2016relation} analyzed the tradeoff between utility and data privacy, and compared the performance of different data privacy metrics, including local differential privacy, identifiability, and mutual information. \blue{It is unclear how effective such data privacy metrics are at protecting inference privacy in a decentralized network. We address this issue in this paper by studying the relationships between data and inference privacy metrics.} 

\blue{The paper \cite{bordenabe2016correlated} analyzed the relationship between privacy leakage and correlation between the private hypothesis and sensor observations. The authors' aim was to recover a public hypothesis correlated with both the private hypothesis and sensor observation. Data privacy was not considered.} The authors of \cite{PinCalmon2012} proposed three inference privacy metrics to measure the exposure of the private hypothesis: information privacy, differential privacy (as applied to the private hypothesis instead of the sensor data \blue{and which we call inference differential privacy in this paper to avoid confusion}), and average information leakage. They showed that information privacy is the strongest among the three, while inference differential privacy does not guarantee information privacy. Methods using the information privacy metric, both nonparametric \cite{SunTayHe2017towards,he2017multi,al2017ratio,HeTaySun:C16,SunTay2017,hamm2017} and parametric \cite{PinCalmon2012}, have been proposed in the literature. Average information leakage is used by \cite{salamatian2013hide} and \cite{Yamamoto1983} to restrict the leakage of sensitive information. The references \cite{he2018latent,SonWanTay:C18} consider the tradeoff between prediction accuracy of sensitive information or parameters and data utility. \blue{These works do not consider the simultaneous protection of both inference and data privacy.}

\blue{Different metrics have been proposed to measure privacy leakage.} The reference \cite{wang2016relation} studied the relationship between various data privacy metrics under a distortion utility but did not consider any inference privacy metrics, whereas \cite{PinCalmon2012} compared only inference privacy metrics. \blue{However, the works mentioned above only compare metrics for inference or data privacy separately. To protect both inference and data privacy, we need to analyze the interplay of the privacy metrics.} Inference privacy and data privacy generally do not imply each other. In \cite{liao2017hypothesis}, maximum leakage is used as the privacy metric to limit inference privacy leakage and the authors conclude that this leads to data privacy leakage. On the other hand, data privacy constraints do not prevent the fusion center from making statistical inference. This is because data privacy metrics do not distinguish between the public and private hypotheses. If the data privacy budget is chosen in such a way that the private hypothesis is difficult to infer, it also means that the utility of inferring the public hypothesis will be severely impacted. A more technical discussion of the relationship between inference and data privacy metrics is provided in \cref{sec:metrics}. 

Several works have considered both inference and data privacy constraints. The paper \cite{chechik2002extracting} proposed an iterative optimization method to protect against average information leakage (inference privacy) and mutual information privacy (data privacy). However, it is unclear if these are the best inference and data privacy metrics for a decentralized IoT network. For a decentralized sensor network, \cite{he2017customized} proposed the use of local differential privacy to achieve both data and inference privacy (which they call inherent and latent privacy, respectively). However, the proposed approach is computationally expensive as it involves a brutal force search. Furthermore, local differential privacy also does not distinguish between the public and private hypotheses of interest. It is thus a ``blunt'' privacy protection approach. In \cite{hamm2017}, the author proposed a two-stage approach, with one stage implementing an inference privacy mechanism, and the other stage a local differential privacy mechanism. In this paper, we adopt a similar two-stage approach. In addition, we study the relationship between possible data and inference privacy metrics, which was not done in \cite{hamm2017}.  

\subsection{Our Contributions}\label{subsec:Contributions}
In this paper, we develop a joint inference and data privacy-preserving framework for a decentralized IoT network \cite{ChaVar:86,Tsi:93,ChaVee:03,TayTsiWin:J08b,Tay:J12,Tay:J15,ZhaChoPez:13,HoTayQue:J15}. Our main contributions are as follows. 
\begin{enumerate}
\item To the best of our knowledge, the interplay between inference privacy and data privacy and the relationship between different privacy metrics have not been adequately investigated. In this paper, we introduce the concept of privacy implication \blue{with vanishing budget}, and show how one privacy metric is related to another in this framework. We argue that in a practical IoT network, both information privacy and local differential privacy metrics should be incorporated in each sensor's privacy mapping to provide suitable inference and data privacy guarantees, respectively. We then propose an optimization framework with joint information privacy and local differential privacy constraints.
\item We propose a local privacy mapping for each sensor that consists of two local privacy mappings concatenated together. One local privacy mapping implements an information privacy mechanism while the other implements a local differential privacy mechanism. We propose two different architectures depending on the order of concatenation. We show that both information privacy and local differential privacy are preserved in post-processing, and local differential privacy is immune to pre-processing, which imply that our proposed architectures achieve the given privacy budgets.  
\end{enumerate}

\blue{Simulations demonstrate that our proposed architectures can protect both information privacy and local differential privacy, while maximizing the detection accuracy of the public hypothesis. To test our proposed joint information privacy and local differential privacy framework, we perform experiments using empirical datasets. However, in these cases, the sensor observations' distributions are unknown \emph{a priori}. Therefore, we adopt an empirical risk optimization framework modified from \cite{SunTayHe2017towards} to now include both information privacy and local differential privacy constraints. Experiments demonstrate that our proposed approach can achieve a good utility-privacy tradeoff.} 

\blue{This paper is an extension of our conference paper \cite{SunTay2017}, which utilized a nonparametric approach to learn sensor decision rules with both local differential privacy and information privacy constraints. In this paper, we rigorously prove the relationships between different privacy metrics under the concept of privacy implication, and propose architectures to achieve both information privacy and local differential privacy in the parametric case. Additional simulations that provide insights into the performance of different architectures as well as experiments on real data sets are also included in this journal version.} 

The rest of this paper is organized as follows. In Section~\ref{sec:formulation}, we present our system model. In Section~\ref{sec:metrics}, we introduce the concept of privacy implication and non-guarantee, review the definition of various privacy metrics, and show the relationships between them. We propose a parametric approach with local differential privacy and information privacy constraints in Section~\ref{sec:parametric}, while a non-parametric approach is discussed in Section~\ref{sec:nonparametric}. Simulation results are shown in Section~\ref{sec:simulation}, and we conclude in Section~\ref{sec:conclusion}.

\emph{Notations:}
We use capital letters like $X$ to denote random variables or vectors, lowercase letters like $x$ for deterministic scalars, and boldface lowercase letters like $\bx$ for deterministic vectors. The vector $\bzero$ has all zero entries, and $\bm{1}$ has all ones. We use $\Gamma^c$ to denote the complement of the set $\Gamma$. We assume that all random variables are defined on the same underlying probability measure space with probability measure $\P$. We use $p_X(\cdot)$ to denote the probability mass function of $X$, and $p_{X\mid Y}(\cdot \mid \cdot)$ to denote the conditional probability mass function of $X$ given $Y$. We use $I(\cdot\ ;\ \cdot)$ to denote mutual information. We use $\log$ to denote natural logarithm, \blue{and $\epsilon_i \downarrow 0$ to mean that the sequence $\epsilon_1,\epsilon_2,\ldots$ decreases to 0.} We say that two vectors $\bx$ and $\bx'$ are neighbors if they differ in only one of their vector components \cite{wang2003using, xiong2016randomized, liao2017hypothesis}, and we denote this by $\bx \sim \bx'$. 

\section{System Model}
\label{sec:formulation}

We consider $s$ sensors making observations generated by a public hypothesis $H$ and a private hypothesis $G$, as shown in \cref{fig:PBL}. Each sensor $t\in\{1,2,\ldots,s\}$, makes a noisy observation $X_t = x_t \in \calX$. Each sensor $t$ then summarizes its observation $X_t=x_t$ using a local decision rule or privacy mapping $p_t: \calX \mapsto \calZ$ and transmits $Z_t=z_t\in\calZ$ to a fusion center with probability $p_t(z_t|x_t)=p_{Z_t|X_t}(z_t|x_t)$. Both $\calX$ and $\calZ$ are assumed to be discrete alphabets. Let $X = (X_t)_{t=1}^s\in\calX^s$ denote the observations of all sensors, and $Z = (Z_t)_{t=1}^s\in\calZ^s$ denote the transmitted information from all sensors. 

The fusion center infers the public hypothesis $H$ from $Z$. However, it can also use $Z$ to infer $G$, even though it has not been authorized to do so. At the same time, it may also try to recover $X$ from $Z$. In this paper, for simplicity, we consider the case where $H\in\{0,1\}$ is a binary hypothesis (our work is easily extended to the multiple hypothesis case), and $G = (G_1,\ldots,G_q) \in \calG = \{0,1\}^q$ is a random vector where each component is binary, i.e., $G$ is a $2^q$-ary hypothesis. Our goal is to design privacy mappings at sensors in order to make it difficult for the fusion center to both infer $G$ (inference privacy) and to recover $X$ (data privacy), while allowing it to infer $H$ with reasonable accuracy. In this paper, we do not make any assumptions regarding the conditional independence of sensor observations, which is common in many of the works in decentralized detection \cite{ChaVar:86,Tsi:93,ChaVee:03,TayTsiWin:J08b,Tay:J12,Tay:J15,ZhaChoPez:13,HoTayQue:J15}.


 In the example of fall detection, whether a fall happens is the public hypothesis $H$. Each binary $G^i$, $i=1,\ldots,q$, in the private hypothesis $G$ can correspond to detecting if the person is performing different activities like running, climbing stairs, squatting, and so on. 

The utility of the network is the probability of inferring $H$ correctly by the fusion center. Inference privacy is measured by the ``difficulty'' of inferring $G$. One of our objectives is to determine which inference privacy metric is most suitable for the IoT network in \cref{fig:PBL}. Furthermore, since some sensors' observations may be uncorrelated with $G$, the raw observations from these sensors are transmitted to the fusion center to maximize the utility. There is then leakage of data privacy for these sensors. Therefore, we also require that the local privacy mappings at each sensor incorporate a data privacy mechanism.

\section{Relationships Between Privacy Metrics}
\label{sec:metrics}
In this section, we consider different privacy metrics proposed in the literature and study their relationships to provide insights into the best inference and data privacy metrics for a decentralized IoT network. A privacy budget $\epsilon\geq 0$ is associated with each type of privacy metric, with a smaller $\epsilon$ corresponding to a more stringent privacy guarantee. We consider the following inference and data privacy metrics. Note that we use the joint distribution $p_{G,X,Z}$ in \cref{inference_privacy,data_privacy} although \cref{inference_privacy} (inference privacy) depends only on $p_{G,Z}$ while \cref{data_privacy} (data privacy) depends only on $p_{X,Z}$. This is done to make it easier to present \cref{def:implication}, which allows us to relate inference and data privacy metrics. 

\begin{Definition}[Inference privacy metrics]\label{inference_privacy} 
Let $\epsilon\geq0$. We say that $p_{G,X,Z}$ satisfies each of the following types of \emph{inference privacy} if the corresponding conditions hold.
\begin{itemize}
    \item $\epsilon$-\blue{inference} differential privacy \cite{PinCalmon2012}: for all $\bg,\bg'\in\calG$ such that $\bg\sim \bg'$, and $\bz\in\calZ^s$, 
		\begin{align*}
		\frac{p_{Z|G}(\bz|\bg)}{p_{Z|G}(\bz|\bg')}\leq e^{\epsilon}.
		\end{align*}
    \item $\epsilon$-average information leakage \cite{PinCalmon2012}: $I(G;Z)\leq\epsilon$.
    \item $\epsilon$-information privacy \cite{PinCalmon2012}: for all $\bg\in\calG$ and $\bz\in\calZ^s$, 
		\begin{align*}
		e^{-\epsilon}\leq\frac{p_{G|Z}(\bg|\bz)}{p_{G}(\bg)}\leq e^{\epsilon}.
		\end{align*}
\end{itemize}
\end{Definition}

\blue{Note that we use the term ``inference differential privacy'' in \cref{inference_privacy} to avoid confusion with ``differential privacy'', which is usually associated with protecting the privacy of the data $X$. In \cref{inference_privacy}, the differential privacy refers to that for the private hypothesis $G$.}

\begin{Definition}[Data privacy metrics]\label{data_privacy}
Let $\epsilon\geq0$. We say that $p_{G,X,Z}$ satisfies each of the following types of \emph{data privacy} if the corresponding conditions hold.
\begin{itemize}
    \item $\epsilon$-local differential privacy \cite{duchi2013local}: for each sensor $t\in\{1,2,\ldots,s\}$, and all $x,x'\in\calX$, and $z\in\calZ$, 
		\begin{align*}
		\frac{p_t(z|x)}{p_t(z|x')}\leq e^{\epsilon}.
		\end{align*}
    \item $\epsilon$-mutual information privacy \cite{wang2016relation}: $I(X;Z)\leq\epsilon$.
    \item $\epsilon$-identifiability \cite{wang2016relation}: for all $\bx,\bx'\in\calX^s$ such that $\bx\sim\bx'$, and $\bz\in\calZ^s$, 
		\begin{align*}
		\frac{p_{X|Z}(\bx|\bz)}{p_{X|Z}(\bx'|\bz)}\leq e^{\epsilon}.
		\end{align*}
\end{itemize}
\end{Definition}

To relate one privacy metric to another, we introduce the concept of privacy implication with vanishing budget in the following definition.

\begin{Definition}[Privacy implication with vanishing budget]\label{def:implication}
We say that Type~A privacy implies Type~B privacy, if for all sequences of probability distributions $(p_{G,X,Z}^{i})_{i\geq 1}$ such that $p_{G,X,Z}^{i}$ satisfies $\epsilon_i$-Type~A privacy with $\epsilon_i\downarrow 0$, then $p_{G,X,Z}^{i}$ satisfies $\epsilon^{\prime}_{i}$-Type~B privacy with $\epsilon^{\prime}_{i}\downarrow 0$.
\end{Definition}

\blue{In nontechnical terms, \cref{def:implication} says that arbitrarily strong Type~A privacy implies arbitrarily strong Type~B privacy. Therefore, to achieve a desired level of Type~B privacy, it suffices to ensure that Type~A privacy with sufficiently small budget is satisfied. Conversely, we say Type~A privacy does not guarantee Type~B privacy if the condition in Definition~\ref{def:implication} does not hold, i.e.,} there exists a sequence of probability distributions $(p_{G,X,Z}^{i})_{i\geq 1}$, such that $p_{G,X,Z}^{i}$ satisfies $\epsilon_i$-Type~A privacy with $\epsilon_i\downarrow 0$, and $\epsilon^\prime_i$-Type~B privacy with $\inf_{i\geq 1} \epsilon^{\prime}_{i}>0$.


\begin{figure*}[!htb]
\centering
\includegraphics[width=0.83\textwidth]{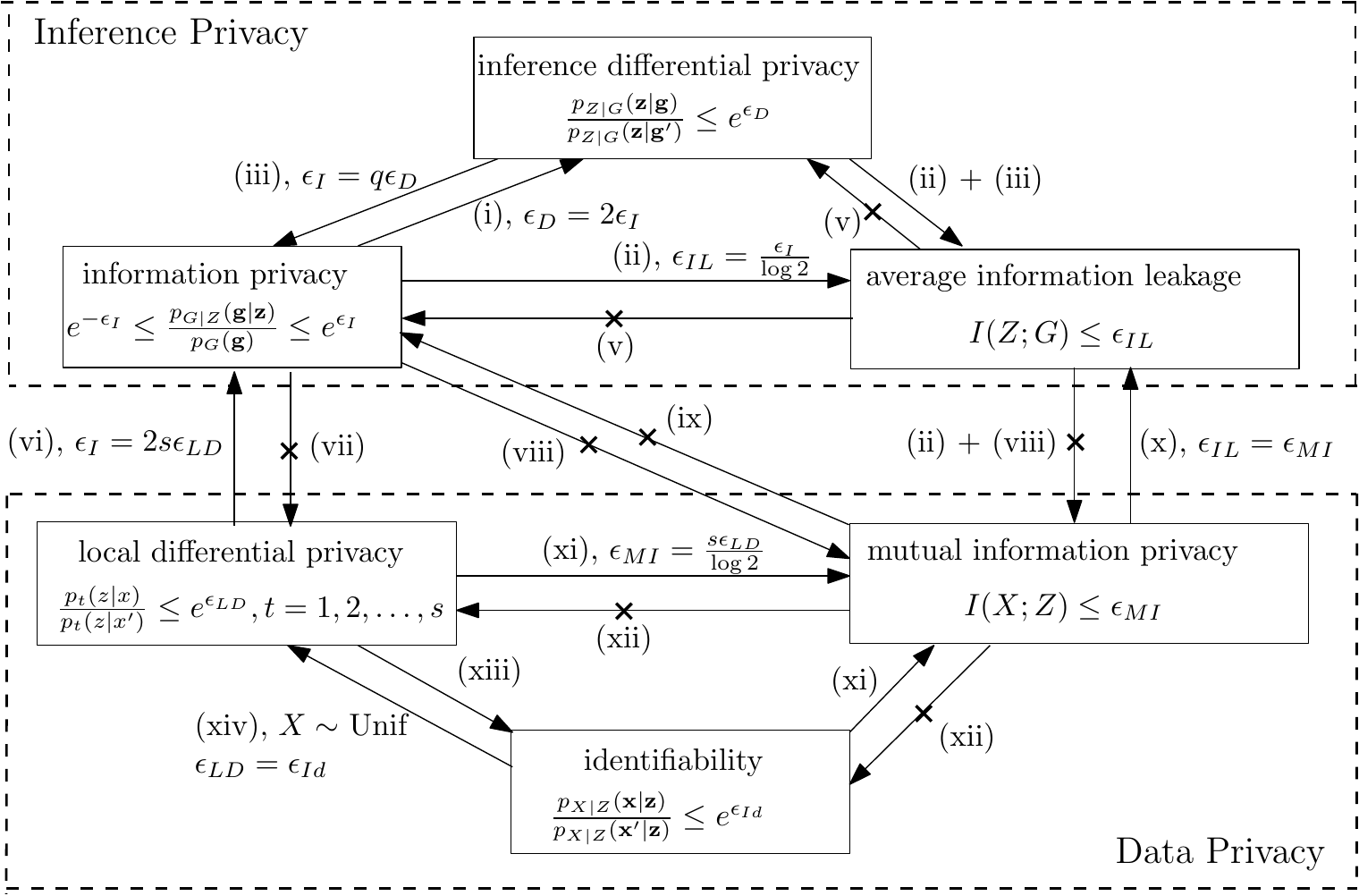}\caption{Relationships between different privacy metrics for an IoT network with fixed number of private hypothesis components $q$ and number of sensors $s$. An arrow $\rightarrow$ means ``implies'' while $\mathrlap{\ \ \times}\longrightarrow$ means ``does not guarantee''.}
\label{fig:Metrics}
\end{figure*}

The following theorem elucidates the relationships between different privacy metrics, which are summarized in Fig.~\ref{fig:Metrics}. Some of these relationships are results proven in \cite{PinCalmon2012}, and are reproduced here for completeness.

\begin{Theorem}\label{thm:privacy_metrics}
Consider the decentralized IoT network in \cref{fig:PBL} with $s\geq 1$ sensors and $G=(G_1,\ldots,G_q)$. Let $\epsilon\geq0$. Then, the following holds for $p_{G,X,Z}$.
\begin{enumerate}[(i)]
    \item \label{it:i2d}\blue{\cite[Theorem 3]{PinCalmon2012}} $\epsilon$-information privacy implies $2\epsilon$-\blue{inference} differential privacy for all $s\geq 1$.
    \item \label{it:i2ai}\blue{\cite[Theorem 3]{PinCalmon2012}} $\epsilon$-information privacy implies $\frac{\epsilon}{\log{2}}$-average information leakage for all $s\geq 1$.
    \item \label{it:d2i} $\epsilon$-\blue{inference} differential privacy implies $q\epsilon$-information privacy. If $q\to\infty$, then \blue{inference} differential privacy does not guarantee information privacy.
		\item \label{it:d2ai} $\epsilon$-\blue{inference} differential privacy implies $\frac{q\epsilon}{\log 2}$-average information leakage. If $q\to\infty$, then \blue{inference} differential privacy does not guarantee average information leakage.
    \item \label{it:ai2i}Average information leakage does not guarantee information privacy and \blue{inference} differential privacy.
    \item \label{it:ld2i} $\epsilon$-local differential privacy implies $2s\epsilon$-information privacy.
    \item \label{it:i2ld}Information privacy does not guarantee local differential privacy.
    \item \label{it:i2mi}Information privacy does not guarantee mutual information privacy.
    \item \label{it:mi2i}Mutual information privacy does not guarantee information privacy.
    \item \label{it:mi2ai}$\epsilon$-mutual information privacy implies $\epsilon$-average information leakage.
    \item \label{it:ld2mi}$\epsilon$-local differential privacy implies $\frac{s\epsilon}{\log 2}$-mutual information privacy.
    \item \label{it:mi2ld}Mutual information privacy does not guarantee local differential privacy.
    \item \label{it:ld2id}$\epsilon$-local differential privacy yields $(\epsilon+\delta_X)$-identifiability, where $\delta_X = \max \log p_X(\bx)/p_X(\bx')$ with the maximum taken over all neighboring $\bx,\bx'\in\calX^s$. Therefore, $\epsilon$-local differential privacy implies $\epsilon$-identifiability if $X$ is restricted to have uniform distribution on $\calX^s$. Otherwise, local differential privacy does not guarantee identifiability. 
    \item \label{it:id2ld} $\epsilon$-identifiability yields $(\epsilon+\delta_X)$-local differential privacy. Therefore, $\epsilon$-identifiability implies $\epsilon$-local differential privacy if $X$ is restricted to have uniform distribution on $\calX^s$. Otherwise, identifiability does not guarantee local differential privacy.
\end{enumerate}
\end{Theorem}

\begin{IEEEproof}
See \cref{prf:thm:privacy_metrics}.
\end{IEEEproof}

From \cref{thm:privacy_metrics}, we see that information privacy implies the other types of inference privacy metrics in \cref{inference_privacy}. 
\blue{Although for a fixed number of components $q$ of the private hypothesis $G=(G_1,\ldots,G_q)$, inference differential privacy also implies other types of inference privacy  metrics including information privacy, it does not guarantee information privacy when $q\to\infty$.

For data privacy, \cref{thm:privacy_metrics} shows that local differential privacy implies mutual information privacy. As the identifiability metric is essentially the same as local differential privacy up to a fixed constant, we consider only the local differential privacy metric in this paper.}

Although local differential privacy implies information privacy for a fixed number $s$ of sensors, this is no longer true if $s$ is not fixed or known in advance. Furthermore, even if $s$ is known \emph{a priori}, \cref{thm:privacy_metrics} suggests that to achieve $\epsilon$-information privacy based solely on preserving local differential privacy, the order of magnitude of the local differential privacy budget has to be not more than $\epsilon/s$. Note that since the definition of local differential privacy does not distinguish between the public hypothesis $H$ or the private hypothesis $G$, this implies that $p_{H,X,Z}$ also satisfies $\epsilon$-information privacy. If $s$ is large, \cite[Theorem 1(i)]{SunTayHe2017towards} then implies that the \blue{Type I and II errors (the probability of rejecting a true null hypothesis and the probability of rejecting a false null hypothesis, respectively)} for detecting the public hypothesis $H$ also become large, which is therefore undesirable. Hence, we propose to design the sensors' privacy mappings using both information privacy and local differential privacy constraints, where the local differential privacy budget can be chosen to be sufficiently large to achieve a reasonable utility for $H$ while maintaining strong information privacy for $G$.

Therefore, in summary, we propose to use information privacy as the metric for inference privacy to protect the private hypothesis $G$, and local differential privacy as the metric for data privacy of $X$. 
In the subsequent sections, we propose frameworks for designing the local privacy mappings for sensors in a decentralized IoT network under both the parametric and nonparametric cases. These privacy mappings are designed to achieve both information privacy and local differential privacy at the fusion center. 

\section{Parametric Case: Concatenated Privacy Mappings}
\label{sec:parametric}

In this section, we consider the parametric case where $p_{X,H,G}$ is known \emph{a priori}. We first study decentralized detection that preserves only data privacy using the local differential privacy metric. Then we include information privacy as an additional constraint to achieve inference privacy, and propose a local privacy mapping consisting of two concatenated privacy mappings that implement information privacy and local differential privacy mechanisms separately. 

\subsection{Data Privacy using Local Differential Privacy}

We first consider the case where local differential privacy is adopted as the privacy metric for the IoT network in \cref{fig:PBL}. Let $\calQ$ denote the set of $p_{Z\mid X}$ such that
\begin{subequations}\label{Q}
\begin{align}
    &p_{Z\mid X}(\bz\mid \bx)=\prod_{t=1}^s p_t(z_t\mid x_t),\\
    &\sum_{z_t\in\calZ} p_t(z_t \mid  x_t) =1,\\
    &p_t(z_t \mid x_t)\geq 0,\ \forall\ x_t \in \calX,\ z_t\in\calZ,\ t=1,\ldots,s.
\end{align}
\end{subequations}
Let $\gamma_H(Z)$ denote the decision rule used by the fusion center to infer the public hypothesis $H$ from the received sensor information $Z$. Our goal is to
\begin{align}\label{ld_privacy}
\begin{aligned}
&\min_{\gamma_H,p_{Z|X}\in\calQ}\P(\gamma_H(Z)\neq H)\\
&\text{s.t. }\frac{p_t(z|x)}{p_t(z|x')}\leq e^{\epsilon_{LD}}, \forall z\in\calZ,x,x'\in\calX,t=1,2,\ldots,s,
\end{aligned}
\end{align}
where $\epsilon_{LD}\geq 0$ is the local differential privacy budget.

\blue{We use the block nonlinear Gauss-Siedel method \cite{Grippo2000} to optimize \eqref{ld_privacy}: to minimize a continuous differentiable function $f(\bx)$ over $\bx\in\calX_1\times\calX_2\times\ldots\times\calX_s$, at each iteration $k\geq1$ and for each index $i=1,\ldots,s$ in sequential order, we find 
\begin{align*}
x^{i,k} = \argmin_{y\in\calX^i} f(x^{1,k},\ldots,x^{i-1,k},y,x^{i+1,k-1},\ldots,x^{s,k-1}).
\end{align*}
The initial estimates $(x^{i,0})_{i=1}^s$ at iteration $k=0$ are chosen randomly.}

\blue{To apply the block nonlinear Gauss-Siedel method to \eqref{ld_privacy}, we iteratively optimize over the random variables.} For fixed $p_t,t=1,2,\ldots,s$, \eqref{ld_privacy} is a convex optimization over $\gamma_H$\cite{Boyd2004}, which can be solved with standard approaches. Then for each $t=1,\ldots,s$, we fix $\gamma_H$ and $p^i$ where $i\ne t$ and optimize for $p_t$. This procedure is then repeated until a convergence criterion is met.

\begin{Theorem}\label{thm:diff}
Suppose $|\calZ|=2$. Consider optimizing \cref{ld_privacy} over $p_t$ with $\gamma_H$ and $p^i$,$i\ne t$ fixed. The optimal solution is
\begin{align}
\begin{aligned}\label{diff_opt}
    p_t(1|x)=
        \begin{cases}
        \ofrac{1+e^{\epsilon_{LD}}},\text{if }f_t(1,x)\geq f_t(2,x),\\
        \frac{e^{\epsilon_{LD}}}{1+e^{\epsilon_{LD}}},\text{if }f_t(1,x)< f_t(2,x),\\
        \end{cases}\\
     p_t(2|x)=
        \begin{cases}
        \frac{e^{\epsilon_{LD}}}{1+e^{\epsilon_{LD}}},\text{if }f_t(1,x)\geq f_t(2,x),\\
        \ofrac{1+e^{\epsilon_{LD}}},\text{if }f_t(1,x)< f_t(2,x),
        \end{cases}
\end{aligned}
\end{align}
where
\begin{align*}
&f_t(z,x)\\
&=\sum_{{\substack{\bz\in\Psi(z)\\ \bx\in \{\bx : x_t=x\}}}}
\prod_{i\neq t}p^i(\bz|\bx)\left(p_{X,H}(\bx,0)-p_{X,H}(\bx,1)\right),
\end{align*}
with $\Psi(z)=\{\bz : \gamma_H(\bz)=1,z_t=z\}$.
\end{Theorem}
\begin{IEEEproof}
Let $\Gamma =\{\bz : \gamma_H(\bz)=1\}$. We have
\begin{align}
&\P(\gamma_H(Z)\neq H)\nonumber\\
&= p_H(1) + \sum_{\bz\in\Gamma}(p_{Z|H}(\bz|0)p_H(0)-p_{Z|H}(\bz|1)p_H(1))\nonumber\\
&=p_H(1)+\sum_{\bz\in\Gamma,\bx} p_{Z|X}(\bz|\bx) \left(p_{X,H}(\bx,0)-p_{X,H}(\bx,1)\right)\nonumber\\
&=p_H(1)+\sum_{z\in\calZ,x\in\calX}p_t(z|x)f_t(z,x)\nonumber\\
&=p_H(1)+\sum_{x\in\calX}p_t(1|x)(f_t(1,x)-f_t(2,x)) + \sum_{x\in\calX} f_t(2,x).\label{p_obj}
\end{align}
We rewrite \cref{ld_privacy} as the following linear programming problem:
\begin{align}
\begin{aligned}\label{ldp_l}
\min_{p_t} \ & \sum_{x\in\calX}p_t(1|x)(f_t(1,x)-f_t(2,x))\\
\text{s.t. }& p_t(z|x)-e^{\epsilon_{LD}}p_t(z|x')\leq 0\\
&p_t(z|x)\geq 0, \sum_{z} p_t(z|x)=1, z\in\calZ,x,x'\in\calX.
\end{aligned}
\end{align}
Without loss of generality, assume $a=p_t(1|1)\geq p_t(1|2)\geq\ldots\geq p_t(1||\calX|)=b$ satisfy the constraints of \eqref{ldp_l}. From \cref{p_obj}, to minimize $\P(\gamma_H(Z)\neq H)$, we have $p_t(1|x)=a$ for $x\in\calX_1=\{x : f_t(2,x)>f_t(1,x)\}$ and $p_t(1|x)=b$ for $x\in\calX_2=\{x : f_t(2,x)\leq f_t(1,x)\}$. Thus, we can simplify \eqref{ldp_l} to
\begin{align*}
\min_{a,b} \ & \sum_{x\in\calX_1} a (f_t(1,x)-f_t(2,x))\\
&+\sum_{x\in\calX_2}b (f_t(1,x)-f_t(2,x))\\
\text{s.t. }& a-e^{-\epsilon_{LD}}b\geq 0\\
& a-e^{\epsilon_{LD}}b\leq 0\\
& (1-a)-e^{-\epsilon_{LD}}(1-b)\geq 0\\
& (1-a)-e^{\epsilon_{LD}}(1-b)\leq 0\\
&a\geq 0,b\geq 0.
\end{align*}
It can be shown that the solution to the above linear program is 
\begin{align*}
a= \frac{e^{\epsilon_{LD}}}{1+e^{\epsilon_{LD}}},\ b= \ofrac{1+e^{\epsilon_{LD}}},
\end{align*}
which proves the theorem.
\end{IEEEproof}

\blue{\Cref{thm:diff} provides a closed form solution for the local differential privacy mapping at each sensor $t$ when the sensor is constrained to be binary. This is typically the case when the sensor is low-cost and has limited computational resources. The result in \cref{thm:diff} thus allows efficient implementation in practice.}

\subsection{Joint Inference and Data Privacy}

\begin{figure}[!htb]
\centering
\includegraphics[width=0.45\textwidth]{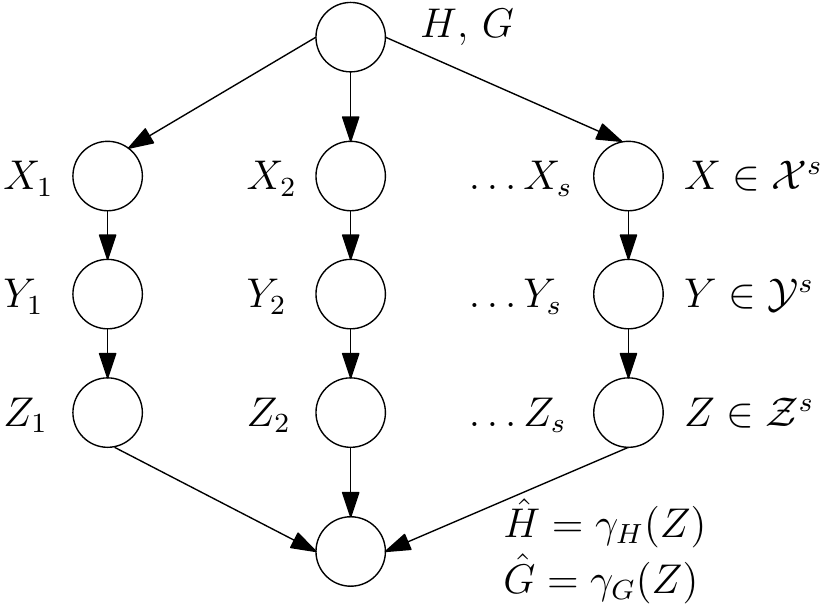}
\caption{Each sensor $t$'s privacy mapping $p_t(z_t|x_t) = p_t^1(y_t|x_t)\cdot p_t^2(z_t|y_t)$ consists of two privacy mappings concatenated together.}
\label{fig:PBL2}
\end{figure}

From \cref{thm:privacy_metrics}, as information privacy is one of the strongest inference privacy metrics, we adopt the information privacy metric when designing our privacy mechanism. To achieve joint inference and data privacy, we consider
\begin{align}\tag{P0}\label{main}
\begin{aligned}
&\min_{\gamma_H,p_{Z|X}\in\calQ}\P(\gamma_H(Z)\neq H),\\
&e^{-\epsilon_I}\leq\frac{p_{Z|G}(\bz|\bg)}{p_{Z}(\bz)}\leq e^{\epsilon_I}, \\
&\hspace{2cm}\forall \bz\in\calZ^s,\ \forall \bg=\{g^1,g^2,\ldots,g^q\}\in\calG,\\
&\frac{p_t(z|x)}{p_t(z|x')}\leq e^{\epsilon_{LD}},\forall z\in\calZ,\ x,x'\in\calX,t=1,2,\ldots,s,
\end{aligned}
\end{align}
where $\epsilon_I>0$ and $\epsilon_{LD}$ are the information privacy budget and local differential privacy budget, respectively.

Since \eqref{main} is a NP-complete problem\cite{TsiAth:85}, we seek to find suboptimal solutions rather than solving \eqref{main} directly. Similar to the work in \cite{hamm2017}, we break the privacy mapping $p_{Z|X}$ in \eqref{main} into two concatenated stages as shown in \cref{fig:PBL2}, where sensor observations $X\in\calX^s$ are first mapped to $Y\in\calY^s$, which is then mapped to $Z\in\calZ^s$, i.e., the mappings $p_{Y|X}$ and $p_{Z|Y}$ satisfy 
\begin{subequations}\label{2stagemapping}
\begin{align}
&p_{Y|X}(\by|\bx)=\prod_{t=1}^s p_t^1(y_t|x_t),\ p_{Z|Y}(\bz|\by) = \prod_{t=1}^s p_t^2(z_t|y_t),\\
&\text{and for all } t=1,\ldots,s,\\
&p_t^1(y_t|x_t) \geq 0,\ \sum_{y_t} p_t^1(y_t|x_t) = 1,\ \forall y_t\in\calY, x_t\in\calX,\\
&p_t^2(z_t|y_t) \geq 0,\ \sum_{z_t} p_t^2(z_t|y_t) = 1,\ \forall z_t\in\calZ, y_t\in\calY.
\end{align}
\end{subequations}
The local privacy mapping for each sensor $t$ is given by
\begin{align}
p_t(z\mid x)= \sum_{y\in\calY} p_t^2(z|y)p_t^1(y|x). 
\end{align}
We propose the following two architectures: 
\begin{enumerate}
	\item Information-LocaL differential privacy (ILL): the mapping from $X\in\calX^s$ to $Y\in\calY^s$ preserves information privacy, while the mapping from $Y\in\calY^s$ to $Z\in\calZ^s$ preserves local differential privacy.
	\item Local differential-Information Privacy (LIP): the mapping from $X\in\calX^s$ to $Y\in\calY^s$ preserves local differential privacy, while the mapping from $Y\in\calY^s$ to $Z\in\calZ^s$ preserves information privacy.
\end{enumerate}

In the following \cref{prop:ILL,prop:LIP}, we show that this two-stage approach achieves joint inference and data privacy. But first, we discuss how to optimize for the privacy mappings in practice. 

In the ILL architecture, we find mappings $p_{Y\mid X}(\by|\bx) = \prod_{t=1}^s p_t^1(y_t|x_t)$ and $p_{Z\mid Y}(\bz|\by) = \prod_{t=1}^s p_t^2(z_t|y_t)$ satisfying
\begin{align}\tag{P1}\label{ILL_opt}
\begin{aligned}
&\min_{\gamma_H,p_{Y|X},p_{Z|Y}}\P(\gamma_H(Z)\neq H)\\ 
&e^{-\epsilon_I}\leq\frac{p_{Y|G}(\by|\bg)}{p_{Y}(\by)}\leq e^{\epsilon_I}, \forall \bg\in\calG,\by\in\calY^s,\\
&\frac{p_t^2(z|y)}{p_t^2(z|y')}\leq e^{\epsilon_{LD}/2},\forall z\in\calZ,\ y,y'\in\calY,t=1,2,\ldots,s,\\
&p_{Y|X}, p_{Z|Y} \text{ satisfy \cref{2stagemapping}.}
\end{aligned}
\end{align}

To solve the problem \cref{ILL_opt}, we first consider the information privacy subproblem:
\begin{subequations}\label{info_sub}
\begin{align}
&\min_{\gamma_H,p_{Y|X}}\P(\gamma_H(Y)\neq H),\\
&e^{-\epsilon_I}\leq\frac{p_{Y|G}(\by|\bg)}{p_{Y}(\by)}\leq e^{\epsilon_I}, \forall \bg\in\calG,\by\in\calY^s, \label{info_sub_const}\\
&p_{Y|X}\text{ satisfy \cref{2stagemapping}.}
\end{align}
\end{subequations}
From \cite[Theorem~2]{SunTayHe2017towards}, to meet the constraint \cref{info_sub_const}, it suffices to ensure that
\begin{align}
\min_{\bg\in\calG\backslash\{\bzero\}, \gamma_G} R_{\bg}(p_{Y\mid X},\gamma_G) \geq \theta,\label{R_const}
\end{align}
where 
\begin{align}
R_{\bg}(p_{Y\mid X},\gamma_G) &= \ofrac{2}\Big(\P(\gamma_{G}(Y)=\bg){G=\bzero}\nonumber\\
&\quad\quad +\P(\gamma_{G}(Y)=\bzero){G=\bg}\Big), 
\end{align}
and $\theta = (1-c_G(1-e^{-\epsilon_I/2}))/2$ with
\begin{align*}
c_G&=\min_{\bg\ne\bzero}\Big \{\P(Y\in\argmin_{\by\in\calY^s}\ell_{\bg}(\by)\mid G=\bzero),\\
&\hspace{1.5cm}\P(Y\in\argmax_{\by\in\calY^s}\ell_{\bg}(\by)\mid G=\bg)\Big \},\\
\ell_{\bg}(\by) &=\frac{p_{Y\mid G}(\by\mid \bg)}{p_{Y\mid G}(\by\mid \bzero)},\\
p_{Y|G}(\by|\bg) &= \sum_{\bx} p_{Y|X}(\by|\bx)p_{X|G}(\bx|\bg)\\
&=\prod_{t=1}^s \sum_{x_t\in\calX} p_t^1(y_t|x_t)p_{X_t|G}(x_t|\bg). 
\end{align*}
By using the constraint \cref{R_const}, we reduce the $2|\calG|\times|\calY|^s$ constraints in \cref{info_sub_const} to a single (but weaker) constraint, which is easier to optimize in practice. A block nonlinear Gauss-Siedel method variant of \eqref{info_sub} similar to that used for solving \cref{ld_privacy} can then be used to find the privacy mapping $p_{Y\mid X}$ \blue{as follows. 
\begin{enumerate}[(i)]
	\item For a fixed privacy mapping $p_{Y|X}$, we first find the optimal fusion center decision rule $\gamma_H$. 
	\item For each sensor $t=1,\ldots,s$ in sequential order, we optimize for sensor $t$'s information privacy mapping $p_t^1(y_t|x_t)$, with $\gamma_H$ and the privacy mappings of all other sensors $p^1_{\backslash t}=\prod_{j\neq t} p_j^1$ fixed. Let the set of sensor $t$'s information privacy mapping be $\Phi$. The optimization is done by solving the following linear program: 
\begin{align*}
\min_{\nu_\phi}\ &\sum_{\phi\in\Phi}\nu_\phi L_H(\phi)\\
\text{s.t.}\ & \sum_{\phi\in\Phi}\nu_\phi \min_{\gamma_G}R_{\bg}(\phi\cdot p^1_{\backslash t},\gamma_{G})\geq\theta,\ \forall \bg\in\calG\backslash\{\bzero\},\\
&\sum_{\phi\in\Phi} \nu_\phi=1,\ \nu_\phi\geq 0,\ \forall\phi\in\Phi.
\end{align*}
where $L_H(\phi)$ is $\P(\gamma_H(Y)\neq H)$ when the privacy mapping $p_{Y|X} = \phi\cdot p^1_{\backslash t}$. Note that from \cite[Section~II.B]{Poor2013}, the decision rule $\gamma_G=\argmin_{\gamma}R_{\bg}(\phi\cdot p^1_{\backslash t},\gamma)$ is given by
\begin{align*}
  \gamma_G(\by)
	=\begin{cases}
    1,\text{ if  } \ell_{\bg}(\by) \geq 1,\\
    0, \text{ otherwise}.
   \end{cases}
\end{align*}
\end{enumerate}  
The above two steps are iterated until a convergence criterion (e.g., when the $L_1$ norm of the difference in the mapping $p_{Y|X}$ between two successive iterations is less than a small constant) is met. 
}

In the second stage, we consider the local differential privacy subproblem:
\begin{align}\label{diff_sub}
\begin{aligned}
&\min_{\gamma_H,p_{Z|Y}}\P(\gamma_H(Z)\neq H),\\ 
&\frac{p_t^2(z|y)}{p_t^2(z|y')}\leq e^{\epsilon_{LD}/2},\forall z\in\calZ,\ y,y'\in\calY,t=1,2,\ldots,s.\\
& p_{Z|Y} \text{ satisfy \cref{2stagemapping}.}
\end{aligned}
\end{align}
If $|\calZ|=2$, the solution follows from \cref{thm:diff}. If $|\calZ| > 2$, we can use a standard linear program solver \cite{Lofberg2004} for \cref{diff_sub} (see the discussion leading to \cref{ldp_l} on how to formulate this linear program).

Similarly, for the LIP architecture, we consider the following optimization problem:
\begin{align}\tag{P2}\label{LIP_opt}
\begin{aligned}
&\min_{\gamma_H,p_{Y|X},p_{Z|Y}}\P(\gamma_H(Z)\neq H),\\ 
&\frac{p_t^1(y|x)}{p_t^1(y|x')}\leq e^{\epsilon_{LD}},\forall y\in\calY,\ x,x'\in\calX,t=1,2,\ldots,s,\\
&e^{-\epsilon_I}\leq\frac{p_{Z|G}(\bz|\bg)}{p_{Z}(\bz)}\leq e^{\epsilon_I}, \forall \bg\in\calG,\bz\in\calZ^s,\\
&p_{Y|X}, p_{Z|Y} \text{ satisfy \cref{2stagemapping}.}
\end{aligned}
\end{align}
Solving \cref{LIP_opt} can be done in a similar fashion as \cref{ILL_opt}. 
 
We next show that the concatenation of information privacy mapping with local differential privacy mapping achieves joint information and local privacy in both the ILL and LIP architectures. 

\begin{Proposition}\label{prop:ILL}
Let $\epsilon_I, \epsilon_{LD} \geq 0$. Suppose that $p_{G,X,Y}$ satisfies $\epsilon_I$-information privacy and $p_{G,Y,Z}$ satisfies $\epsilon_{LD}/2$-local differential privacy. Then, the following holds.
\begin{enumerate}[(a)]
	\item For any randomized mapping $p_{Z\mid Y}$, $p_{G,X,Z}$ satisfies $\epsilon_I$-information privacy.
	\item For any randomized mapping $p_{Y\mid X}$, $p_{G,X,Z}$ satisfies $\epsilon_{LD}$-local differential privacy.
\end{enumerate}
\end{Proposition}
\begin{IEEEproof}
\begin{enumerate}[(a)]
	\item For any $\bz\in\calZ^s$ and $\bg\in\calG$, we have
	\begin{align*}
	\frac{p_{Z|G}(\bz|\bg)}{p_{Z}(\bz)}=\frac{\sum_{\by} p_{Z|Y}(\bz|\by)p_{Y|G}(\by|\bg)}{\sum_{\by} p_{Z|Y}(\bz|\by)p_{Y}(\by)}.
	\end{align*}
	Since $e^{-\epsilon_I}\leq\frac{p_{Y|G}(\by|\bg)}{p_{Y}(\by)}\leq e^{\epsilon_I}$ for all $\by\in\calY^s$, we obtain $e^{-\epsilon_I}\leq\frac{p_{Z|G}(\bz|\bg)}{p_{Z}(\bz)}\leq e^{\epsilon_I}$.
	
	\item Consider any sensor $t$. For any $y,y'\in\calY$ and $z\in\calZ$, we have $e^{-\epsilon_{LD}/2}\leq\frac{p_t^2(z|y)}{p_t^2(z|y')}\leq e^{\epsilon_{LD}/2}$. Therefore, for any $x,x'\in\calX$, we then have
\begin{align*}
\frac{p_t(z|x)}{p_t(z|x')}
&=\frac{\sum_{y} p_t^2(z|y)p_t^1(y|x)}{\sum_{y} p_t^2(z|y)p_t^1(y|x')}\\
&\leq\frac{\sum_{y} e^{\epsilon_{LD} /2}p_t^2(z|y')p_t^1(y|x)}{\sum_{y} e^{-\epsilon_{LD} /2}p_t^2(z|y')p_t^1(y|x')}\\
&=e^{\epsilon_{LD}},
\end{align*}
for a fixed $y'\in\calY$.
\end{enumerate}
The proposition is now proved.
\end{IEEEproof}
\Cref{prop:ILL} shows that joint information privacy for $G$ and local differential privacy for $X$ are preserved in the ILL architecture. In the LIP architecture, it is clear that information privacy for $G$ is preserved since this is an explicit constraint in \cref{LIP_opt}. Local differential privacy preservation follows from \cite[Proposition 2.1]{dwork2014algorithmic}, which is reproduced below for completeness.

\begin{Proposition}\label{prop:LIP}
Let $\epsilon_{LD} \geq 0$. Suppose that $p_{G,X,Y}$ satisfies $\epsilon_{LD}$-local differential privacy. Then for any randomized mapping $p_{Z\mid Y}$, $p_{G,X,Z}$ satisfies $\epsilon_{LD}$-local differential privacy.
\end{Proposition}
\begin{IEEEproof}
For any sensor $t$, $z\in\calZ$, $x,x'\in\calX$, we have
\begin{align*}
\frac{p_t(z|x)}{p_t(z|x')}
&=\frac{\sum_{y} p_t^2(z|y)p_t^1(y|x)}{\sum_{y} p_t^2(z|y)p_t^1(y|x')}\leq e^{\epsilon_{LD}},
\end{align*}
since $\frac{p_t^1(y|x)}{p_t^1(y|x')}\leq e^{\epsilon_{LD}}$. The proposition is now proved.
\end{IEEEproof}

\section{Nonparametric Case: Empirical Risk Optimization}\label{sec:nonparametric}

In many IoT applications, knowing the joint distribution of $(H,G)$ and the sensor observations is impractical due to difficulties in accurately modeling this distribution. \blue{To overcome this, we can adopt a nonparametric approach similar to the NPO framework in \cite{SunTayHe2017towards} to convert \cref{main} into an empirical risk optimization approach. NPO in \cite{SunTayHe2017towards} finds a privacy mapping that satisfies an information privacy constraint. To adapt to \cref{main}, we can simply add the additional linear constraints corresponding to local differential privacy to that framework. For the full details, we refer the reader to \cite{SunTayHe2017towards} and the supplementary material in the final part of this paper. For convenience, we call this approach the} Empirical information and local differential PrIvaCy (EPIC) optimization.

\section{Numerical Results}\label{sec:simulation}
In this section, we carry out simulations and experiments on real datasets to verify the performance of the proposed optimization framework using joint information privacy and local differential privacy constraints.

\subsection{Parametric Case Study}

We first consider the performance of ILL and LIP in \cref{sec:parametric}. In our simulations, we consider binary public hypothesis $H$ and private hypothesis $G$. To evaluate the performance, we compute the Bayes probability errors for detecting $H$ and $G$ since these are the minimum detection errors any detector can achieve so that our results are oblivious to the choice of learning method adopted by the fusion center. The Bayes error of detecting $H$ reflects the utility of our method, while the Bayes error of detecting $G$ reflects the inference privacy of the private hypothesis $G$. Data privacy of the sensor $t$'s observation $X_t$ is quantified by the mutual information $I(X_t;Z_t)$.

\begin{figure}[!htb]
  \centering
  \includegraphics[width=0.45\textwidth]{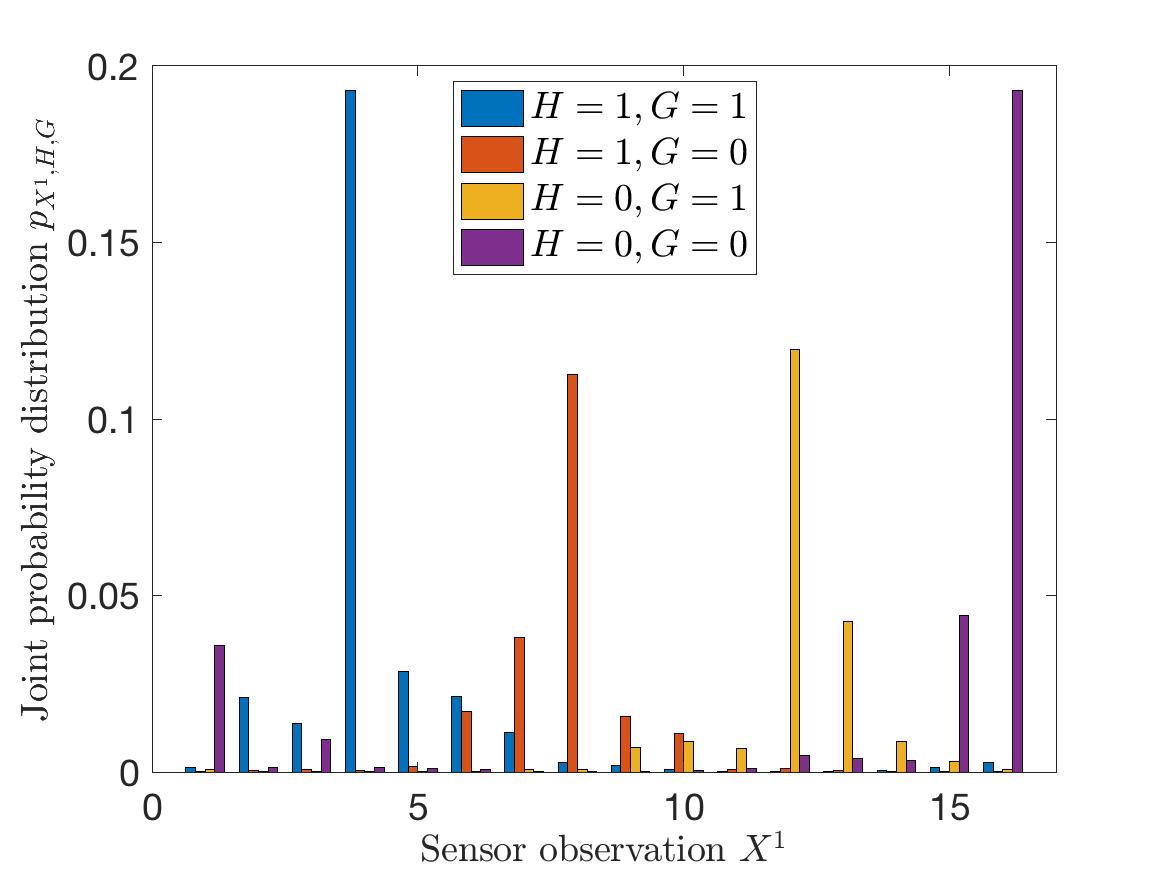}
  \caption{Joint distribution $p_{X_1, H, G}$ of sensor observation, public hypothesis $H$ and private hypothesis $ G$. The correlation coefficient between $H$ and $ G$ is $0.2$. }
  \label{fig:distribution}
\end{figure}

Consider a network of $6$ sensors and a fusion center. Suppose that $\calX=\{1,2,\ldots,16\}$ and $\calZ=\{1,2\}$. We set the correlation coefficient between the public hypothesis $H$ and private hypothesis $G$ to be $0.2$. We assume that each sensor has identical joint distribution as shown in \cref{fig:distribution}.

\begin{figure}[!htb]
\centering
\includegraphics[width=0.9\linewidth]{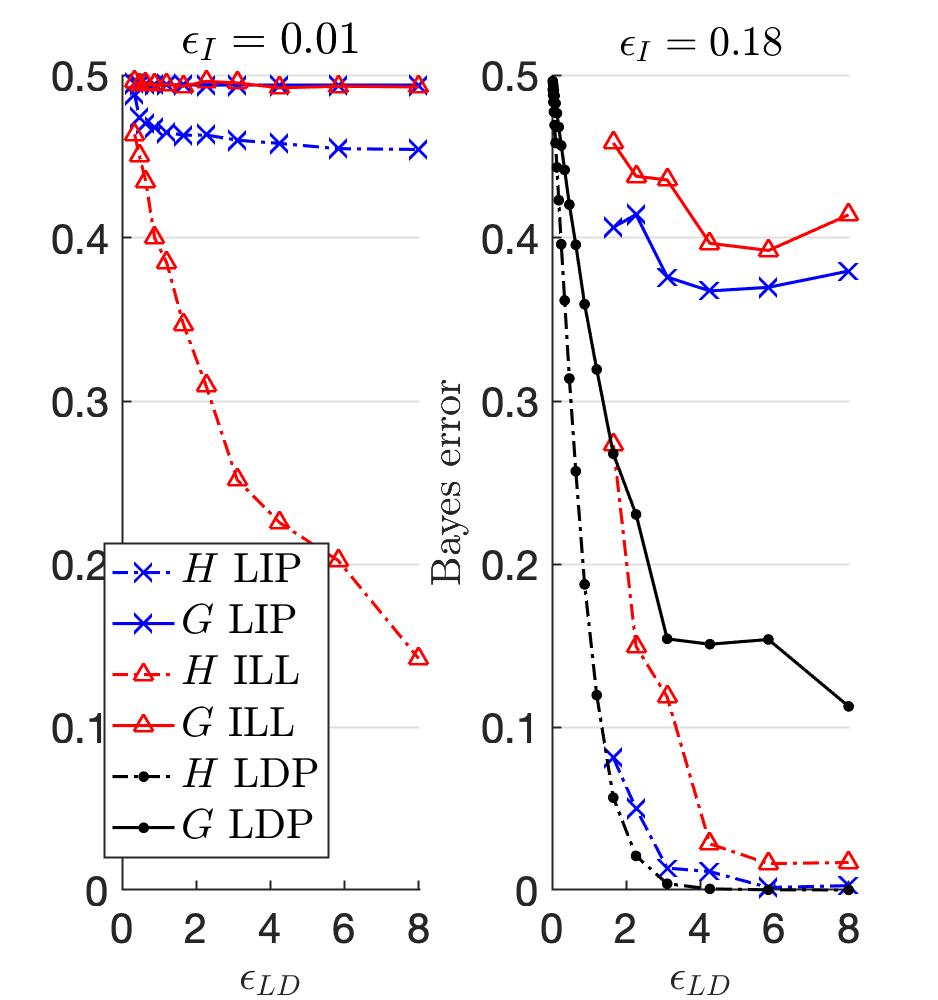}
\caption{Bayes error for detecting $H$ and $G$ under LIP and ILL for fixed privacy threshold ratio $r$ and varying local differential privacy budget $\epsilon_{LD}$.}
\label{fig:info_fixed}
\end{figure}

In \cref{fig:info_fixed}, we let the information privacy budget be fixed at $\epsilon_I=0.01$ and $0.18$, and vary the local differential privacy budget $\epsilon_{LD}$. We see that if $\epsilon_I$ is small, ILL is better at inferring the public hypothesis $H$ while achieving a similar detection error for the private hypothesis $G$ when compared to LIP. This is because ILL first sanitizes the sensor observations $X$ for information privacy before applying a local differential privacy mapping, which allows it better control over sanitization of statistical information needed to infer $G$ but keeping information for inferring $H$. On the other hand, if $\epsilon_I$ is large, LIP infers $H$ with better accuracy. We also compare with the approach that uses only a local differential privacy constraint (i.e., the information privacy constraint in \cref{main} is removed), which we call LDP in the left drawing in \cref{fig:info_fixed}. Without any constraint on $\epsilon_I$, we see that LDP gives poor information privacy protection for $G$.

\begin{figure}[!htb]
\centering
\includegraphics[width=0.9\linewidth]{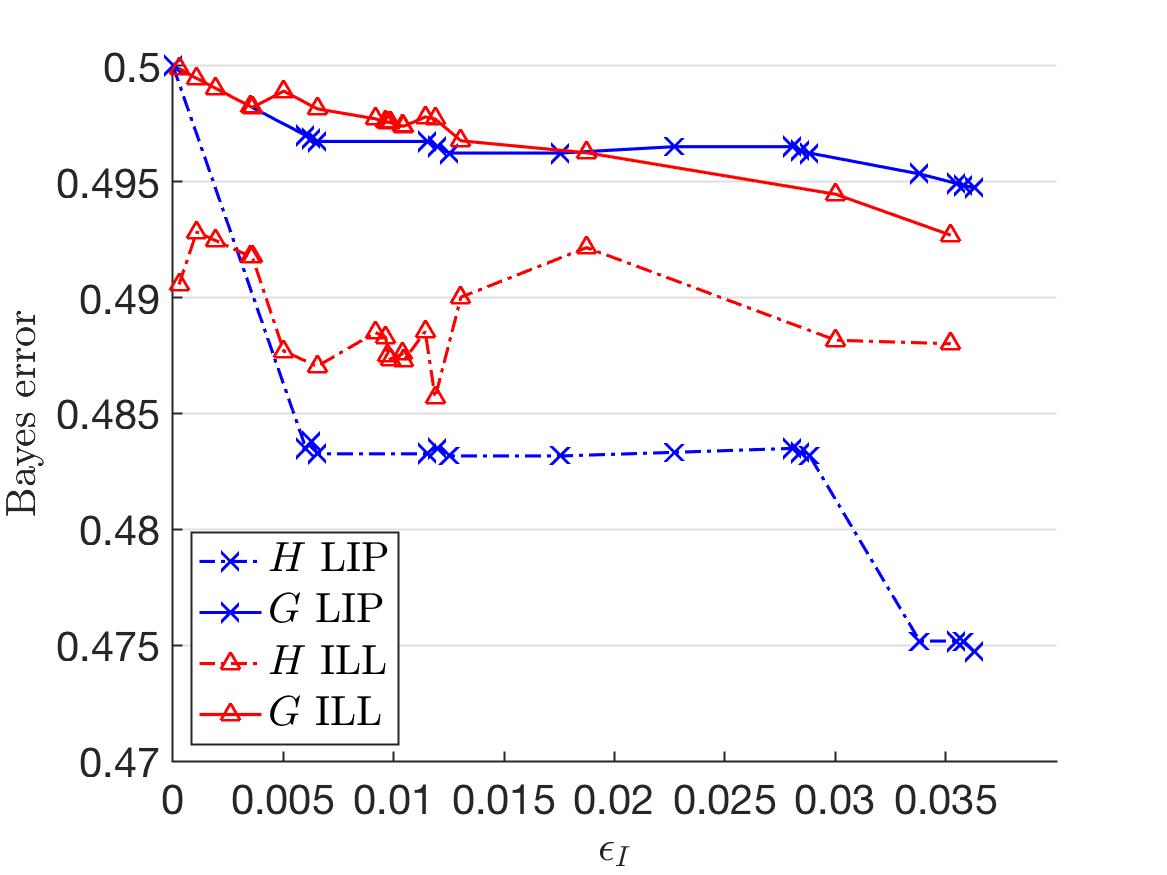}
\caption{Bayes error for detecting $H$ and $G$ under LIP and ILL for fixed local differential privacy budget $\epsilon_{LD}$ and varying $\epsilon_I$.}
\label{fig:diff_fixed}
\end{figure}

In \cref{fig:diff_fixed}, we fix $\epsilon_{LD}=0.07$, while varying $\epsilon_I$. We see that when $\epsilon_{LD}$ is small, the Bayes error of detecting $H$ is large regardless of the value of $\epsilon_I$. This aligns with our discussion after \cref{thm:privacy_metrics} that we should not use local differential privacy to achieve inference privacy for the private hypothesis $G$ as this approach also leads to a poor inference performance for the public hypothesis $H$.

\begin{figure}
    \centering
    \begin{subfigure}[b]{0.4\textwidth}
        \includegraphics[width=0.9\textwidth]{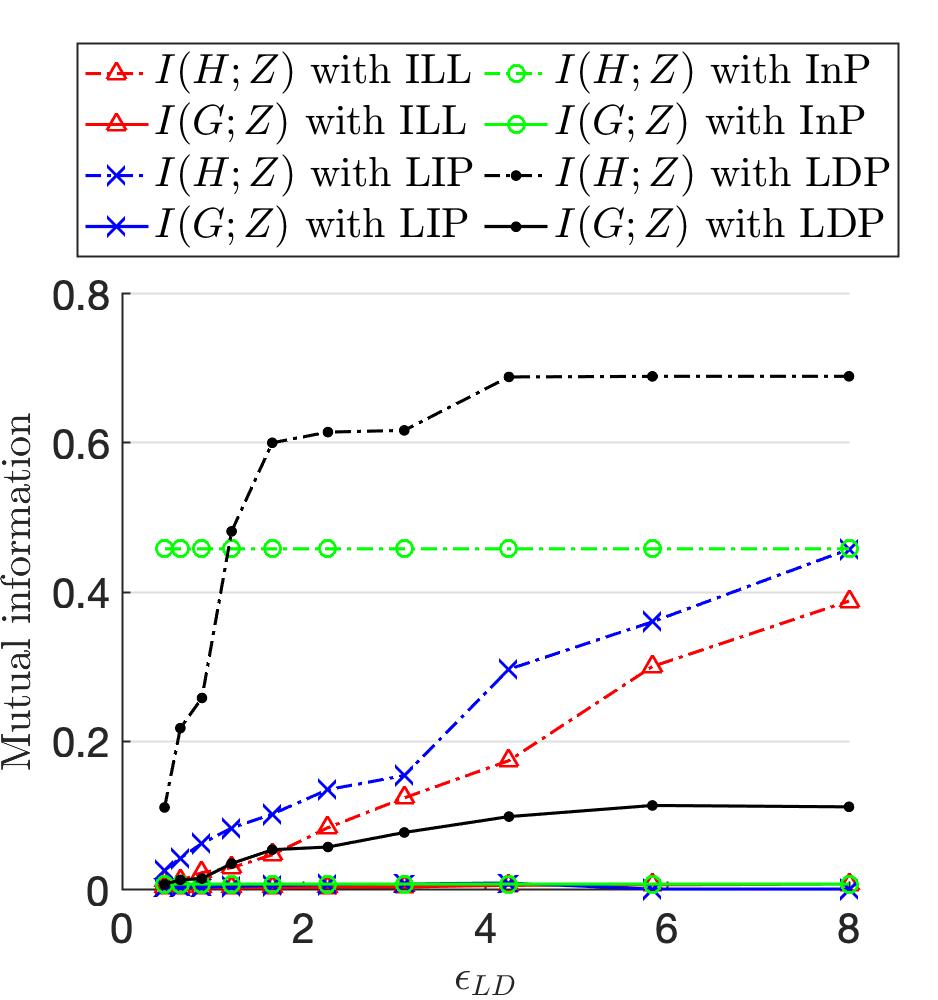}
        \caption{Mutual information $I(H;Z)$ and $(G;Z)$.}
        \label{fig:mutual_hg}
    \end{subfigure}
    \begin{subfigure}[b]{0.4\textwidth}
        \includegraphics[width=0.9\textwidth]{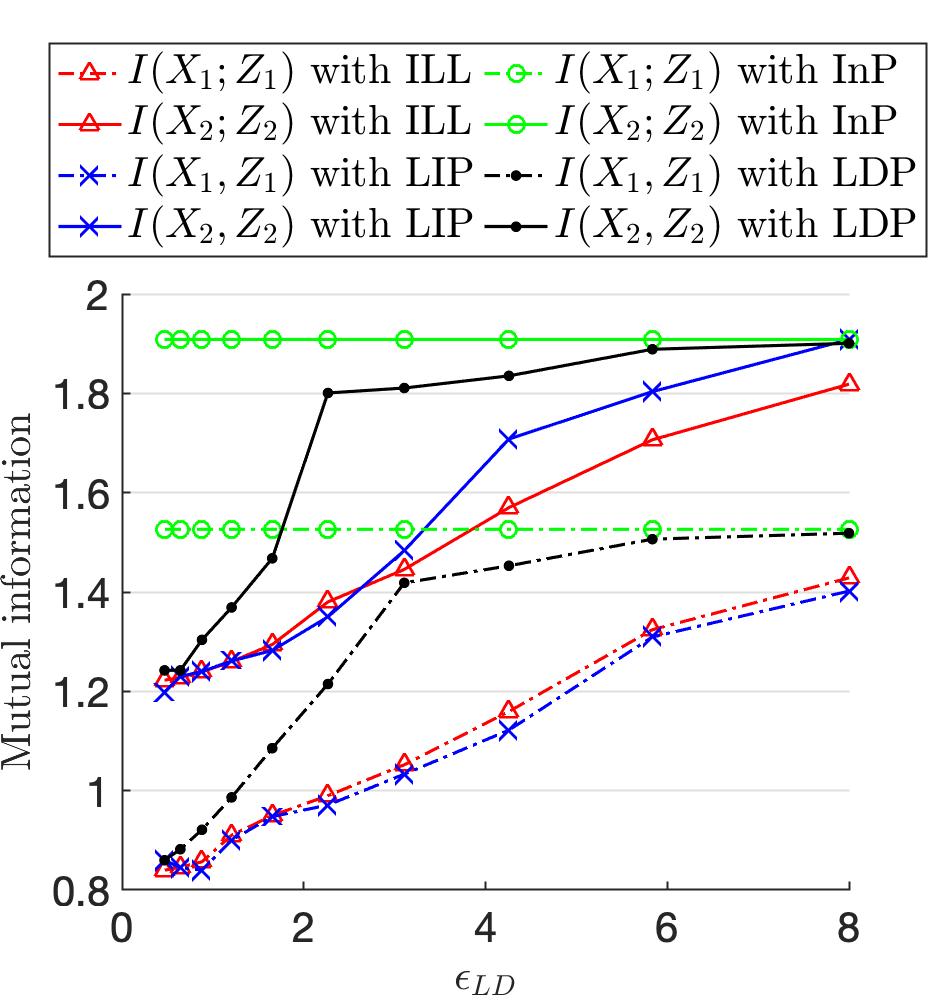}
        \caption{Mutual information $I(X_1;Z_1)$ and $(X_2;Z_2)$.}
        \label{fig:mutual_13}
    \end{subfigure}
    \caption{Mutual information with $\epsilon_I=0.15$ and varying $\epsilon_{LD}$ for ILL, LIP, InP and LDP.}
    \label{fig:mutual}
\end{figure}

We next consider the case where sensor 1's observations are independent of $G$ with marginal conditional distribution under $H$ same as the joint distribution shown in \cref{fig:distribution}. All other sensors follow the distribution in \cref{fig:distribution}. In \cref{fig:mutual}, we fix $\epsilon_I=0.15$ and vary $\epsilon_{LD}$ to illustrate the mutual information between different quantities. We also compare with the approach that uses only an information privacy constraint (i.e., the local differential privacy constraint in \cref{main} is removed), which we call InP. From Fig.~\ref{fig:mutual_hg}, we observe that both ILL and LIP yield sanitized information $Z$ that have a high mutual information with the public hypothesis $H$, and low mutual information with the private hypothesis $G$. However, with LDP the mutual information  $I(H;Z)$ and $I(G;Z)$ are both much higher compared to other methods, since it does not protect the information privacy of $G$.

In Fig.~\ref{fig:mutual_13}, we compare the mutual informations $I(X_1;Z_1)$ and $I(X_2,Z_2)$ under different privacy architectures. We see that $I(X_1;Z_1)$ under ILL and LIP are much lower than that under InP. In particular, InP does not achieve good data privacy for $X_1$ since the information privacy constraint only removes statistical information in $X_1$ related to $G$, which in this case is none as $X_1$ is independent of $G$. This example illustrates the need to include both inference and data privacy constraints in our privacy mapping design. We also see that $I(X_2 ; Z_2)$ under both ILL and LIP is lower than that under InP, but converges to that of InP as $\epsilon_{LD}$ becomes bigger.

\subsection{Nonparametric Case Study: OPPORTUNITY Data Set and Adult Data Set}

We test the nonparametric EPIC framework in \cref{sec:nonparametric} on the OPPORTUNITY Activity Recognition Data Set \cite{Chavarriaga2013} \blue{and the Adult Data Set \cite{kohavi1996scaling}} available at UCI Repository\cite{Lichman:2013}, and compare its performance with RUCA\cite{al2017ratio}, DCA\cite{kung2017discriminant} and MDR\cite{diamantaras2016data}. In EPIC, we set the local decision space of each sensor to be $\calZ=\{1,2\}$.

\subsubsection{Data Preprocessing}
In the OPPORTUNITY Activity Recognition Data Set, measurements from motion sensors including on-body sensors, sensors attached to objects, and ambient sensors like switches, are recorded while a person performs a series of typical daily activities. In this experiment, our public hypothesis $H$ is whether the person is standing or walking, while the private hypothesis $G$ is whether the person is touch a drawer or dishwasher. We used data from the `S2-Drill' dataset, and sklearn \cite{scikit-learn} to select $s=15$ sensors that are the most correlated with our chosen labels. Since the sensor reading is continuous, unsupervised discretization was applied to quantize each continuous sensor reading to $10$ levels. We randomly sampled $n=80$ instances of training data, and $3427$ instances of testing data.

\blue{
In the Adult Data Set, basic information of a certain population such as age, work class, education, income, marriage status was collected. In our experiment, we set the public hypothesis to be whether a person's income is greater than \$$50,000$ or not. The private hypothesis is the 3-ary hypothesis that the person is married (denoted as `Married-civ-spouse’, 'Married-spouse-absent’ and `Married-AFspouse’ in the data set), used to be married (denoted as `Separated’, `Divorced' and `Widowed’ in the data set) and Never married ('Never-married' in the data set). We select age, workclass, education-num, race, sex as the features, which represents the sensor observation $X$ in our problem formulation. Although the data is not collected from a sensor network, we can still apply our method to this data set. We discretize continuous data to 5 bins and perform one-hot encoding to categorical data. We select $n=120$ instances of training data where both the public and private hypotheses are evenly distributed and $15,050$ instances of testing data.
}

\subsubsection{Comparison Benchmarks}

As comparison benchmarks, we compare our method to the following methods: 
\begin{enumerate}[(i)]
	\item NPO \cite{SunTayHe2017towards}, which is a nonparametric method that considers only information privacy and no data privacy; and
	\item Empirical LDP (E-LDP), which is solving \eqref{objective} without \eqref{maryconstraints}, i.e.,  only local differential privacy is considered.
	\item The centralized approaches RUCA\cite{al2017ratio}, DCA\cite{kung2017discriminant} and MDR\cite{diamantaras2016data}, which require that all sensors send their observations to a central data curator that then applies an overall privacy mapping. Note that since the mapping in RUCA, DCA and MDR are deterministic, they do not provide any local differential privacy protection. 
	\item Sensors do not apply any privacy mapping and send their raw observations to the fusion center, i.e., $Z=X$. In this case, no local differential privacy protection is available, while some information privacy maybe possible depending on the underlying distribution $p_{X\mid G}$. This serves as a benchmark to show the intrinsic error probabilities achievable. 
\end{enumerate}
Similar to \cite{SunTayHe2017towards}, to estimate the privacy budgets achieved by each method, we compute
\begin{align}
\hat{\epsilon}_I &= \max\limits_{g\in\calG,\bz\in\calZ^s} \left|\log \frac{\hat{p}_{G,Z}(g,\bz)}{\hat{p}_G(g)\hat{p}_Z(\bz)}\right|,\\
\hat{\epsilon}_{LD}&=\max_{z\in\calZ,x,x'\in\calX,t\in\{1,\ldots,s\}}\log \frac{p_t(z\mid x)}{p_t(z\mid x')}
\end{align}
as estimates for the information privacy and local differential privacy budgets respectively. Here, $\hat{p}_{A}(a)$ is the empirical probability of the event $\{A=a\}$. Note that a smaller $\hat{\epsilon}$ implies stronger information privacy and a smaller $\hat{\epsilon}_{LD}$ implies stronger local differential privacy. We see that $\hat{\epsilon}_{LD}=\infty$ for RUCA, MDR, and the case $Z=X$.

\subsubsection{Result and Discussion}

From Tables~\ref{tab:activity data set} and \ref{tab:census}, we observe that EPIC achieves the lowest information privacy and local differential privacy budgets compared to all the other benchmarks while maintaining utility similar to the other methods. Compared to NPO, it has similar information privacy budget but significantly lower local differential privacy budget since NPO does not consider any data privacy constraints. It is interesting that EPIC allows further sanitization of the sensor information in order provide data privacy without significantly deteriorating the detection performance of $H$. \blue{Compared to E-LDP, it has similar local differential privacy budget, but a significantly lower information privacy constraint.} Due to having both information privacy and local differential privacy constraints, we see that EPIC has the highest error rate for detecting $H$ amongst all the methods, which is the price it pays for having the least privacy leakage. However, the error rates for $H$ are still within $0.01$ (1\%) of the best error rate amongst the other competing sanitization methods other than $Z=X$. 

\begin{table}[!ht]
\centering
\caption{Detection errors using the OPPORTUNITY Activity Recognition Data Set.}
\begin{tabular}{|c|c|c|c|c|}
\hline
Detection Method&\begin{tabular}[c]{@{}c@{}}$H$\end{tabular} & \begin{tabular}[c]{@{}c@{}}$G$\end{tabular}&$\hat{\epsilon}_I$ &$\hat{\epsilon}_{LD}$\\ \hline
EPIC ($r=0.99,\epsilon_{LD}=1$)                 & $10.91\%$       & $43.65\%$  &$0.46$&$0.81$\\ \hline
NPO ($r=0.99$)                & $10.53\%$       & $43.17\%$  &$0.47$&$2.22$\\ \hline
E-LDP ($\epsilon_{LD}=1$)                & $10.09\%$       & $7.31\%$   &$8.58$&$0.91$\\ \hline
MDR                & $12.56\%$       & $40.16\%$ &  $1.02$ &\multirow{6}{*}{$\infty$}  \\ \cline{1-4}
DCA                & $10.88\%$       & $42.62\%$ &   $0.88$   &  \\ \cline{1-4}
RUCA ($\rho_p=1$)     & $10.23\%$       & $45.73\%$ &  $0.67$  &   \\ \cline{1-4}
RUCA ($\rho_p=100$)     & $10.10\%$       & $43.01\%$ &  $0.69$   &  \\ \cline{1-4}
RUCA ($\rho_p=1000$)     & $10.10\%$       & $43.78\%$ & $0.69$    &  \\ \cline{1-4}
$Z=X$ &$10.05\%$ & $5.57\%$ &$9.14$& \\ \hline
\end{tabular}
\label{tab:activity data set}
\end{table}

\begin{table}[!ht]
\centering
\caption{Detection errors using the Adult Data Set.}
\begin{tabular}{|c|c|c|c|c|}
\hline
Detection Method&\begin{tabular}[c]{@{}c@{}}$H$\end{tabular} & \begin{tabular}[c]{@{}c@{}}$G$\end{tabular}&$\hat{\epsilon}_I$ &$\hat{\epsilon}_{LD}$\\ \hline
EPIC ($r=0.99,\epsilon_{LD}=1$)                 & $37.69\%$       & $62.25\%$  &$0.79$&$0.93$\\ \hline
NPO ($r=0.99$)                & $37.67\%$       & $62.14\%$   &$0.82$&$2.36$\\ \hline
E-LDP ($\epsilon_{LD}=1$)                & $36.11\%$       & $32.91\%$   &$9.48$&$0.97$\\ \hline
MDR                & $37.57\%$       & $64.02\%$     &  $1.68$  &\multirow{6}{*}{$\infty$}  \\ \cline{1-4}
DCA                & $38.38\%$       & $56.33\%$     &  $2.47$ &  \\ \cline{1-4}
RUCA ($\rho_p=1$)     & $41.24\%$       & $65.25\%$     &   $1.61$  &   \\ \cline{1-4}
RUCA ($\rho_p=100$)     & $41.10\%$       & $64.14\%$   &  $1.66$    &  \\ \cline{1-4}
RUCA ($\rho_p=1000$)      & $40.67\%$       & $65.86\%$  &   $1.61$   &  \\ \cline{1-4}
$Z=X$ &$34.05\%$ & $30.48\%$ &$15.33$& \\ \hline
\end{tabular}
\label{tab:census}
\end{table}

\section{Conclusion}\label{sec:conclusion}

We have introduced the concept of privacy implication and non-guarantee to study the relationships between different inference and data privacy metrics. We showed that information privacy and local differential privacy are some of the strongest inference privacy and data privacy metrics, respectively. We considered the problem of preserving both information privacy of a private hypothesis and data privacy of the sensor observations in a decentralized network consisting of multiple sensors and a fusion center, whose task is to infer a public hypothesis of interest. In the parametric case, we proposed two different privacy mapping architectures, and showed that both achieve information privacy and local differential privacy to within the predefined budgets. In the nonparametric case, we proposed an empirical privacy optimization approach to learn the privacy mappings from a given training set. Simulations and tests on real data suggest that our proposed approaches achieve a good utility while protecting both inference and data privacy.

In this paper, we have considered only sensor observations from a single time instance. An interesting future research direction is to generalize our approach to sensor observations over multiple time instances in a dynamic system model.

\begin{appendices}

\section{Proof of Theorem \ref{thm:privacy_metrics}}\label[appendix]{prf:thm:privacy_metrics}

To show privacy non-guarantee, it suffices to provide an example of a sequence of joint distributions not satisfying \cref{def:implication}. We first present such an example that parts of the proof of \cref{thm:privacy_metrics} utilize.

\begin{Example}\label{eg:ai2i}
If the random variables $U\in\calU$ and $V\in\calV$ satisfy the joint distribution as shown in Table~\ref{tab:ai2i}, then we have
\begin{align}
\lim_{\alpha\to 0} I(V;U)&=\lim_{\alpha\to 0} \Big\{ p_{V,U}(0,0)\log\frac{p_{V,U}(0,0)}{p_{V}(0)p_{U}(0)}\\
&\quad +\sum_{i,j\ne0}p_{V,U}(i,j)\log\frac{p_{V,U}(i,j)}{p_{V}(i)p_{U}(j)}\Big\}\nonumber\\
&=\lim_{\alpha\to 0}\left\{ \alpha\log\ofrac{\alpha}+(1-\alpha)\log\ofrac{1-\alpha}\right\}\label{eq:mutualinfo}\\
&=0,\nonumber
\end{align}
and
\begin{align}
\lim_{\alpha\to 0} \frac{p_{V,U}(0,0)}{p_{V}(0)p_{U}(0)}=\lim_{\alpha\to 0} \ofrac{\alpha}=\infty,\label{eq:info}\\
\max_{v,u,u'} \frac{p_{V|U}(v|u)}{p_{V|U}(v|u')}= \frac{p_{V|U}(0|0)}{p_{V|U}(0|1)}=\infty.\label{eq:differential}
\end{align}

\begin{table}[!ht]
\centering
\caption{The joint distribution of $V$ and $U$, where $\alpha\in [0,1]$ is a parameter.}\label{tab:ai2i}
\begin{tabular}{|c|c|c|c|c|c|}
\hline
\multicolumn{2}{|c|}{\multirow{2}{*}{$p_{V,U}$}} & \multicolumn{4}{c|}{$V$}                                                     \\ \cline{3-6}
\multicolumn{2}{|c|}{}                           & $0$      & $1$              & $\ldots$              & $|\calV|-1$            \\ \hline
\multirow{4}{*}{$U$}          & $0$              & $\alpha$ & $0$              & $0$                   & $0$                    \\ \cline{2-6}
                              & $1$              & $0$      & \multicolumn{3}{c|}{\multirow{3}{*}{$\frac{1-\alpha}{(|\calU|-1)(|\calV|-1)}$}} \\ \cline{2-3}
                              & $\vdots$         & $0$      & \multicolumn{3}{c|}{}                                             \\ \cline{2-3}
                              & $|\calU|-1$               & $0$      & \multicolumn{3}{c|}{}                                             \\ \hline
\end{tabular}
\end{table}
\end{Example}

We now proceed with the proof of \cref{thm:privacy_metrics}.
\begin{enumerate}[leftmargin=*, widest=xiii-xiv]
%
    \item[(i-ii)] These claims follow from \cite[Theorem 3]{PinCalmon2012}.
    \item[(iii-iv)] For a fixed $q$, since $p_{G,X,Z}$ satisfies $\epsilon$-\blue{inference} differential privacy, for any $\bz\in\calZ^s,\bg\in\calG$, we have 
		\begin{align*}
		\frac{p_{Z|G}(\bz|\bg_1)}{p_{Z|G}(\bz|\bg_2)}\leq e^{q\epsilon},
		\end{align*}
		for any $\bg_1,\bg_2\in\calG$. Therefore, we have
		\begin{align*}
		e^{-q\epsilon}\leq\frac{p_{G|Z}(\bg|\bz)}{p_G(\bg)}=\frac{p_{Z|G}(\bz|\bg)}{\sum_{\bg'}p_{Z|G}(\bz|\bg')p_{G}(\bg')}\leq e^{q\epsilon}.
		\end{align*}
   Thus $p_{G,X,Z}$ satisfies $q\epsilon$-information privacy. Together with \ref{it:i2ai}, we obtain that $p_{G,X,Z}$ satisfies $q\epsilon/\log 2$-average information leakage.
\ \\
\ \\
If $q\to\infty$, \cite[Theorem 4]{PinCalmon2012} gives an example that shows \blue{inference} differential privacy does not guarantee average information leakage. Together with \ref{it:i2ai}, it implies that \blue{inference} differential privacy does not guarantee information privacy.

    \item[(v)] Substitute $G$ for $U$ and $Z$ for $V$ in Example~\ref{eg:ai2i}, then we get from \eqref{eq:mutualinfo} and \eqref{eq:info}, that average information leakage does not guarantee information privacy. From \ref{it:d2i}, we also obtain that average information leakage does not guarantee \blue{inference} differential privacy.
    \item[(vi)] Since $p_{G,X,Z}$ satisfies $\epsilon$-local differential privacy, for any $\bx_0, \bx\in\calX^s$, and $\bz\in\calZ^s$, we have
		\begin{align*}
		e^{-s\epsilon}p_{Z|X}(\bz|\bx_0)\leq p_{Z|X}(\bz|\bx)\leq e^{s\epsilon}p_{Z|X}(\bz|\bx_0).
		\end{align*}
		Then for any $\bg_1,\bg_2\in\calG,\bz\in\calZ^s$, we have
		\begin{align*}
		&\frac{p_{Z|G}(\bz|\bg_1)}{p_{Z|G}(\bz|\bg_2)}\\
		&=\frac{\sum_{\bx\in\calX^s}p_{Z|X}(\bz|\bx)p_{X|G}(\bx|\bg_1)}{\sum_{\bx\in\calX^s}p_{Z|X}(\bz|\bx)p_{X|G}(\bx|\bg_2)}\\
		&\leq\frac{\sum_{\bx\in\calX^s}e^{s\epsilon}p_{Z|X}(\bz|\bx_0)p_{X|G}(\bx|\bg_1)}{\sum_{\bx\in\calX^s}e^{-s\epsilon}p_{Z|X}(\bz|\bx_0)p_{X|G}(\bx|\bg_2)}\\
		&=e^{2s\epsilon},
		\end{align*}
		from which we obtain 
		\begin{align*}
		e^{-2s\epsilon}\leq\frac{p_{G|Z}(\bg|\bz)}{p_G(\bg)}=\frac{p_{Z|G}(\bz|\bg)}{\sum_{\bg'}p_{Z|G}(\bz|\bg')p_{G}(\bg')}\leq e^{2s\epsilon},
		\end{align*}
		any $\bg\in\calG$ and $\bz\in\calZ$.
				
    \item[(vii-viii)] \label{it:info2mutual}Suppose for any $\bx\in\calX^s,\bg\in\calG$, $p_{X|G}(\bx|\bg)=\ofrac{|\calX|^s}$. Then, $p_{X|G}(\bx|\bg)/p_{X}(\bx)=1$, and
		\begin{align*}
		\frac{p_{Z|G}(\bz|\bg)}{p_{Z}(\bz)}=\frac{\sum_{\bx}p_{Z|X}(\bz|\bx)p_{X|G}(\bx|\bg)}{\sum_{\bx}p_{Z|X}(\bz|\bx)p_{X}(\bx)}=1,
		\end{align*}
		for all privacy mappings $p_{Z|X}$. Therefore, $p_{G,X,Z}$ satisfies $0$-information privacy but does not guarantee local differential privacy and mutual information privacy as $p_{Z|X}$ can be chosen arbitrarily.
    \item[(ix)]   Substitute $X$ for $U$ and $Z$ for $V$ in Example~\ref{eg:ai2i}. From \eqref{eq:mutualinfo}, there is a sequence of distributions $(p^\alpha_{G,X,Z})_{\alpha\geq 0}$ satisfying $\epsilon_\alpha$-mutual information privacy with $\epsilon_\alpha \to 0$ as $\alpha\to 0$. Choose a $\bg_0\in\calG$, and let 
\begin{align*}
p_{X|G}(\bx|\bg_0)=
\begin{cases}
\alpha, & \text{if } \bx=\bzero\\
(1-\alpha)/(|\calX|^s-1), &\text{otherwise}.
\end{cases}
\end{align*} 
For other $\bg\in\calG$, with $\bg\neq\bg_0$, we let $p_{X|G}(\bx|\bg)=\ofrac{|\calX|^s}$ for all $\bx\in\calX^s$. We also let $G$ to be uniformly distributed. Then, we have 
\begin{align*}
p_{Z|G}(\bzero|\bg_0) &= \sum_{\bx} p_{Z|X}(\bzero|\bx) p_{X|G}(\bx|\bg_0) \\
&=p_{Z|X}(\bzero|\bzero) p_{X|G}(\bzero|\bg_0) \\
&=\alpha,
\end{align*}
since $p_{Z|X}(\bzero|\bzero)=1$ from Example~\ref{eg:ai2i}. For $\bg\ne \bg_0$, we also have
\begin{align*}
p_{Z|G}(\bzero|\bg)&=\sum_{\bx} p_{Z|X}(\bzero|\bx) p_{X|G}(\bx|\bg)\\
&=1/|\calX|^s,
\end{align*}
and
\begin{align*} 
p_Z(\bzero)&=\frac{\alpha+\frac{|\calG|-1}{|\calX|^s}}{|\calG|}.
\end{align*}
Therefore, we have 
\begin{align*}
\lim_{\alpha\to 0}\frac{p_{Z|G}(\bzero|\bg_0)}{p_{Z}(\bzero)}=\frac{\alpha |\calG|}{\alpha+\frac{|\calG|-1}{|\calX|^s}}=0,
\end{align*}
which indicates that $\epsilon_I\to\infty$ as $\alpha\to 0$. This proves that mutual information privacy does not guarantee information privacy.

		\item[(x)] Since $I(G;Z|X)=0$, we have $0\leq I(Z;G)=I(X;Z)-I(X;Z|G)\leq I(X;Z)$, and the claim follows immediately.
    \item[(xi)] If $p_{G,X,Z}$ satisfies $\epsilon$-local differential privacy, then $\frac{p_{Z|X}(\bz|\bx_1)}{p_{Z|X}(\bz|\bx_2)}\leq e^{s\epsilon}$, for any $\bx_1,\bx_2\in\calX^s$. The proof then proceeds similarly as that for \ref{it:d2ai}.
    \item[(xii)] Substitute $X$ for $U$ and $Z$ for $V$ in Example~\ref{eg:ai2i}. From \eqref{eq:mutualinfo} and \eqref{eq:differential}, we conclude that mutual information privacy does not guarantee local differential privacy.
    \item[(xiii-xiv)] These claims follow since for any $\bz\in\calZ^s$, $\bx\sim\bx' \in \calX^s$, we have 
		\begin{align*}
		\frac{p_{Z|X}(\bz|\bx)p_{X}(\bx)}{p_{Z|X}(\bz|\bx')p_{X}(\bx')}=\frac{p_{X|Z}(\bx|\bz)}{p_{X|Z}(\bx'|\bz)}.
		\end{align*}
\end{enumerate}
The proof of the theorem is now complete.

\end{appendices}

\bibliographystyle{IEEEtran}

\begin{thebibliography}{10}
\providecommand{\url}[1]{#1}
\csname url@samestyle\endcsname
\providecommand{\newblock}{\relax}
\providecommand{\bibinfo}[2]{#2}
\providecommand{\BIBentrySTDinterwordspacing}{\spaceskip=0pt\relax}
\providecommand{\BIBentryALTinterwordstretchfactor}{4}
\providecommand{\BIBentryALTinterwordspacing}{\spaceskip=\fontdimen2\font plus
\BIBentryALTinterwordstretchfactor\fontdimen3\font minus
  \fontdimen4\font\relax}
\providecommand{\BIBforeignlanguage}[2]{{%
\expandafter\ifx\csname l@#1\endcsname\relax
\typeout{** WARNING: IEEEtran.bst: No hyphenation pattern has been}%
\typeout{** loaded for the language `#1'. Using the pattern for}%
\typeout{** the default language instead.}%
\else
\language=\csname l@#1\endcsname
\fi
#2}}
\providecommand{\BIBdecl}{\relax}
\BIBdecl

\bibitem{roman2013features}
R.~Roman, J.~Zhou, and J.~Lopez, ``On the features and challenges of security
  and privacy in distributed {I}nternet of {T}hings,'' \emph{Computer
  Networks}, vol.~57, no.~10, pp. 2266--2279, Jul. 2013.

\bibitem{SunTay:C16}
M.~Sun and W.~P. Tay, ``Privacy-preserving nonparametric decentralized
  detection,'' in \emph{Proc. IEEE Int. Conf. Acoustics, Speech, and Signal
  Processing}, Shanghai, 2016, pp. 6270--6274.

\bibitem{HeTaySun:C16}
X.~He, W.~P. Tay, and M.~Sun, ``Privacy-aware decentralized detection using
  linear precoding,'' in \emph{Proc. IEEE Sensor Array and Multichannel Signal
  Processing Workshop}, Rio de Janeiro, 2016, pp. 1--5.

\bibitem{he2017multi}
X.~He and W.~P. Tay, ``Multilayer sensor network for information privacy,'' in
  \emph{Proc. IEEE Int. Conf. Acoustics, Speech, and Signal Processing}, New
  Orleans, LA, 2017.

\bibitem{Alemdar2010}
H.~Alemdar and C.~Ersoy, ``Wireless sensor networks for healthcare: A survey,''
  \emph{Computer Networks}, vol.~54, no.~15, pp. 2688--2710, Oct. 2010.

\bibitem{Butun2014}
I.~Butun, S.~D. Morgera, and R.~Sankar, ``A survey of intrusion detection
  systems in wireless sensor networks,'' \emph{{IEEE} Commun. Surveys Tuts.},
  vol.~16, no.~1, pp. 266--282, Jan. 2014.

\bibitem{gdpr}
\BIBentryALTinterwordspacing
(2018) General data protection regulation ({GDPR}). [Online]. Available:
  \url{https://www.eugdpr.org/}
\BIBentrySTDinterwordspacing

\bibitem{dataprotectionact2012}
``Personal data protection act (no 26/2012),'' Republic of Singapore Government
  Gazette, 2012.

\bibitem{digital_privacy_act}
``Digital privacy act (s.c. 2015, c. 32),'' Canada Gazette, 2015.

\bibitem{apple_user_guide}
\BIBentryALTinterwordspacing
(2016) {iPhone} user guide for ios 10. [Online]. Available:
  \url{https://help.apple.com/iphone/10}
\BIBentrySTDinterwordspacing

\bibitem{liu2016dependence}
C.~Liu, S.~Chakraborty, and P.~Mittal, ``Dependence makes you vulnberable:
  Differential privacy under dependent tuples,'' in \emph{Proc. the Network and
  Distributed Sys. Security Symp.}, vol.~16, California, 2016, pp. 21--24.

\bibitem{cormode2011personal}
G.~Cormode, ``Personal privacy vs population privacy: Learning to attack
  anonymization,'' in \emph{Proc. {ACM SIGKDD} Int. Conf. on Knowledge
  Discovery and Data Mining}, California, 2011, pp. 1253--1261.

\bibitem{huang2012differentially}
Z.~Huang, S.~Mitra, and G.~Dullerud, ``Differentially private iterative
  synchronous consensus,'' in \emph{Proc. ACM workshop on Privacy in the
  Electronic Society}, Raleigh, NC, 2012, pp. 81--90.

\bibitem{nozari2015differentially}
E.~Nozari, P.~Tallapragada, and J.~Cort{\'e}s, ``Differentially private average
  consensus with optimal noise selection,'' in \emph{Proc. IFAC Workshop
  Distrib. Estimation Control Networked Syst.}, vol.~48, no.~22, Philadelphia,
  PA, 2015, pp. 203--208.

\bibitem{manitara2013privacy}
N.~E. Manitara and C.~N. Hadjicostis, ``Privacy-preserving asymptotic average
  consensus,'' in \emph{Proc. Eur. Control Conf.}, Zurich, Switzerland, 2013,
  pp. 760--765.

\bibitem{braca2016learning}
P.~Braca, R.~Lazzeretti, S.~Marano, and V.~Matta, ``Learning with privacy in
  consensus $+ $ obfuscation,'' \emph{{IEEE} Trans. Signal Process.}, vol.~23,
  no.~9, pp. 1174--1178, 2016.

\bibitem{mo2017privacy}
Y.~Mo and R.~M. Murray, ``Privacy preserving average consensus,'' \emph{IEEE
  Transactions on Automatic Control}, vol.~62, no.~2, pp. 753--765, 2017.

\bibitem{lazzeretti2014secure}
R.~Lazzeretti, S.~Horn, P.~Braca, and P.~Willett, ``Secure multi-party
  consensus gossip algorithms,'' in \emph{Proc. IEEE Int. Conf. Acoustics,
  Speech, and Signal Processing}, Florence, 2014, pp. 7406--7410.

\bibitem{ambrosin2017odin}
M.~Ambrosin, P.~Braca, M.~Conti, and R.~Lazzeretti, ``Odin: O bfuscation-based
  privacy-preserving consensus algorithm for d ecentralized i nformation fusion
  in smart device n etworks,'' \emph{ACM Trans. on Internet Technology},
  vol.~18, no.~1, p.~6, 2017.

\bibitem{hallgren2017privatepool}
P.~Hallgren, C.~Orlandi, and A.~Sabelfeld, ``Privatepool: privacy-preserving
  ridesharing,'' in \emph{Proc. IEEE Computer Security Found. Symp.}, Santa
  Barbara, CA, 2017, pp. 276--291.

\bibitem{Gentry2009}
C.~Gentry, ``Fully homomorphic encryption using ideal lattices.'' in
  \emph{Proc. ACM Symp. Theory of Comput.}, Bethesda, MD, 2009, pp. 169--178.

\bibitem{brakerski2014leveled}
Z.~Brakerski, C.~Gentry, and V.~Vaikuntanathan, ``({Leveled}) fully homomorphic
  encryption without bootstrapping,'' \emph{ACM Trans. Computation Theory},
  vol.~6, no.~3, p.~13, 2014.

\bibitem{wang2003using}
Y.~Wang, X.~Wu, and H.~Donghui, ``Using randomized response for differential
  privacy preserving data collection,'' in \emph{Proc. {ACM SIGKDD} Int. Conf.
  on Knowledge Discovery and Data Mining}, Washington, D.C., 2003, pp.
  505--510.

\bibitem{xiong2016randomized}
S.~Xiong, A.~D. Sarwate, and N.~B. Mandayam, ``Randomized requantization with
  local differential privacy,'' in \emph{Proc. IEEE Int. Conf. Acoustics,
  Speech, and Signal Processing}, Shanghai, 2016, pp. 2189--2193.

\bibitem{liao2017hypothesis}
J.~Liao, L.~Sankar, F.~P. Calmon, and V.~Y. Tan, ``Hypothesis testing under
  maximal leakage privacy constraints,'' in \emph{Proc. IEEE Int. Symp. on
  Inform. Theory}, Aachen, Germany, 2017, pp. 779--783.

\bibitem{duchi2013local}
J.~C. Duchi, M.~I. Jordan, and M.~J. Wainwright, ``Local privacy and
  statistical minimax rates,'' in \emph{Proc. IEEE Symp. on Foundations of
  Computer Science}, Berkeley, 2013, pp. 429--438.

\bibitem{wang2016relation}
W.~Wang, L.~Ying, and J.~Zhang, ``On the relation between identifiability,
  differential privacy, and mutual-information privacy,'' \emph{{IEEE} Trans.
  Inf. Theory}, vol.~62, no.~9, pp. 5018--5029, Jun. 2016.

\bibitem{bordenabe2016correlated}
N.~E. Bordenabe and G.~Smith, ``Correlated secrets in quantitative information
  flow,'' in \emph{Proc. IEEE Computer Security Found. Symp.}, Lisboa, 2016,
  pp. 93--104.

\bibitem{PinCalmon2012}
F.~du~Pin~Calmon and N.~Fawaz, ``Privacy against statistical inference,'' in
  \emph{Proc. Allerton Conf. on Commun., Control and Computing}, Monticello,
  IL, 2012, pp. 1401--1408.

\bibitem{SunTayHe2017towards}
M.~Sun, W.~P. Tay, and X.~He, ``Toward information privacy for the {Internet of
  Things}: A nonparametric learning approach,'' \emph{{IEEE} Trans. Signal
  Process.}, vol.~66, no.~7, pp. 1734--1747, April 2018.

\bibitem{al2017ratio}
M.~Al, S.~Wan, and S.~Kung, ``Ratio utility and cost analysis for privacy
  preserving subspace projection,'' \emph{arXiv preprint arXiv:1702.07976},
  2017.

\bibitem{SunTay2017}
M.~Sun and W.~P. Tay, ``Inference and data privacy in {IoT} networks,'' in
  \emph{Proc. IEEE Workshop on Signal Processing Advances in Wireless Commun.},
  2017.

\bibitem{hamm2017}
J.~Hamm, ``Enhancing utility and privacy with noisy minimax filters,'' in
  \emph{Proc. IEEE Int. Conf. Acoustics, Speech, and Signal Processing}, New
  Orleans, LA, Mar. 2017, pp. 6389--6393.

\bibitem{salamatian2013hide}
S.~Salamatian, A.~Zhang, F.~du~Pin~Calmon, S.~Bhamidipati, N.~Fawaz, B.~Kveton,
  P.~Oliveira, and N.~Taft, ``How to hide the elephant-or the donkey-in the
  room: Practical privacy against statistical inference for large data,'' in
  \emph{Proc. IEEE Global Conf. on Signal and Information Processing}, Austin,
  TX, 2013, pp. 269--272.

\bibitem{Yamamoto1983}
H.~Yamamoto, ``A source coding problem for sources with additional outputs to
  keep secret from the receiver or wiretappers,'' \emph{{IEEE} Trans. Inf.
  Theory}, vol.~29, no.~6, pp. 918--923, 1983.

\bibitem{he2018latent}
Z.~He, Z.~Cai, and J.~Yu, ``Latent-data privacy preserving with customized data
  utility for social network data,'' \emph{IEEE Transactions on Vehicular
  Technology}, vol.~67, no.~1, pp. 665--673, 2018.

\bibitem{SonWanTay:C18}
Y.~Song, C.~X. Wang, and W.~P. Tay, ``Privacy-aware {Kalman} filtering,'' in
  \emph{Proc. IEEE Int. Conf. Acoustics, Speech, and Signal Processing},
  Calgary, Canada, Apr. 2018.

\bibitem{chechik2002extracting}
G.~Chechik and N.~Tishby, ``Extracting relevant structures with side
  information,'' in \emph{Advances in Neural Information Processing Systems},
  vol.~15.\hskip 1em plus 0.5em minus 0.4em\relax {MIT} Press, 2003, pp.
  881--888.

\bibitem{he2017customized}
Z.~He, Z.~Cai, Y.~Sun, Y.~Li, and X.~Cheng, ``Customized privacy preserving for
  inherent data and latent data,'' \emph{Personal and Ubiquitous Computing},
  vol.~21, no.~1, pp. 43--54, 2017.

\bibitem{ChaVar:86}
Z.~Chair and P.~K. Varshney, ``Optimal data fusion in multiple sensor detection
  systems,'' \emph{{IEEE} Trans. Aerosp. Electron. Syst.}, vol.~22, no.~1, pp.
  98--101, 1986.

\bibitem{Tsi:93}
J.~N. Tsitsiklis, ``Decentralized detection,'' \emph{Advances in Statistical
  Signal Processing}, vol.~2, pp. 297--344, 1993.

\bibitem{ChaVee:03}
J.-F. Chamberland and V.~V. Veeravalli, ``Decentralized detection in sensor
  networks,'' \emph{{IEEE} Trans. Signal Process.}, vol.~51, no.~2, pp.
  407--416, Feb. 2003.

\bibitem{TayTsiWin:J08b}
W.~P. Tay, J.~N. Tsitsiklis, and M.~Z. Win, ``Data fusion trees for detection:
  Does architecture matter?'' \emph{{IEEE} Trans. Inf. Theory}, vol.~54, no.~9,
  pp. 4155--4168, Sep. 2008.

\bibitem{Tay:J12}
W.~P. Tay, ``The value of feedback in decentralized detection,'' \emph{{IEEE}
  Trans. Inf. Theory}, vol.~58, no.~12, pp. 7226--7239, Dec. 2012.

\bibitem{Tay:J15}
------, ``Whose opinion to follow in multihypothesis social learning? {A} large
  deviations perspective,'' \emph{{IEEE} J. Sel. Topics Signal Process.},
  vol.~9, no.~2, pp. 344--359, Mar. 2015.

\bibitem{ZhaChoPez:13}
Z.~Zhang, E.~Chong, A.~Pezeshki, W.~Moran, and S.~Howard, ``Learning in
  hierarchical social networks,'' \emph{{IEEE} J. Sel. Topics Signal Process.},
  vol.~7, no.~2, pp. 305--317, Apr. 2013.

\bibitem{HoTayQue:J15}
J.~Ho, W.~P. Tay, T.~Q. Quek, and E.~K. Chong, ``Robust decentralized detection
  and social learning in tandem networks,'' \emph{{IEEE} Trans. Signal
  Process.}, vol.~63, no.~19, pp. 5019--5032, Oct. 2015.

\bibitem{Grippo2000}
L.~Grippo and M.~Sciandrone, ``On the convergence of the block nonlinear
  {G}auss-{S}eidel method under convex constraints,'' \emph{Operations Research
  Letters}, vol.~26, no.~3, pp. 127--136, 2000.

\bibitem{Boyd2004}
S.~Boyd and L.~Vandenberghe, \emph{Convex Optimization}.\hskip 1em plus 0.5em
  minus 0.4em\relax Cambridge University Press, 2004.

\bibitem{TsiAth:85}
J.~N. Tsitsiklis and M.~Athans, ``On the complexity of decentralized decision
  making and detection problems,'' \emph{{IEEE} Trans. Autom. Control},
  vol.~30, pp. 440--446, 1985.

\bibitem{Poor2013}
H.~V. Poor, \emph{An introduction to signal detection and estimation}.\hskip
  1em plus 0.5em minus 0.4em\relax Springer Science \& Business Media, 2013.

\bibitem{Lofberg2004}
J.~Lofberg, ``Yalmip : A toolbox for modeling and optimization in matlab,'' in
  \emph{Proc. Int. Symp. Computer-Aided Control System Design}, Taipei, 2004,
  pp. 284--289.

\bibitem{dwork2014algorithmic}
C.~Dwork and A.~Roth, ``The algorithmic foundations of differential privacy,''
  \emph{Foundations and Trends{\textregistered} in Theoretical Computer
  Science}, vol.~9, no. 3--4, pp. 211--407, 2014.

\bibitem{Chavarriaga2013}
R.~Chavarriaga, H.~Sagha, A.~Calatroni, S.~T. Digumarti, G.~Tr{\"o}ster, J.~del
  R.~Mill{\'a}n, and D.~Roggen, ``The opportunity challenge: A benchmark
  database for on-body sensor-based activity recognition,'' \emph{Pattern
  Recognition Lett.}, vol.~34, no.~15, pp. 2033--2042, 2013.

\bibitem{kohavi1996scaling}
R.~Kohavi, ``Scaling up the accuracy of naive-bayes classifiers: A
  decision-tree hybrid.'' in \emph{Proc. {ACM SIGKDD} Int. Conf. on Knowledge
  Discovery and Data Mining}, vol.~96, Portland, Oregon, 1996, pp. 202--207.

\bibitem{Lichman:2013}
\BIBentryALTinterwordspacing
M.~Lichman, ``{UCI} machine learning repository,'' 2013. [Online]. Available:
  \url{http://archive.ics.uci.edu/ml}
\BIBentrySTDinterwordspacing

\bibitem{kung2017discriminant}
S.-Y. Kung, ``Discriminant component analysis for privacy protection and
  visualization of big data,'' \emph{Multimedia Tools and Applications},
  vol.~76, no.~3, pp. 3999--4034, 2017.

\bibitem{diamantaras2016data}
K.~Diamantaras and S.~Kung, ``Data privacy protection by kernel subspace
  projection and generalized eigenvalue decomposition,'' in \emph{IEEE Int.
  Workshop Machine Learning for Signal Processing}, Salerno, 2016, pp. 1--6.

\bibitem{scikit-learn}
F.~Pedregosa, G.~Varoquaux, A.~Gramfort, V.~Michel, B.~Thirion, O.~Grisel,
  M.~Blondel, P.~Prettenhofer, R.~Weiss, V.~Dubourg, J.~Vanderplas, A.~Passos,
  D.~Cournapeau, M.~Brucher, M.~Perrot, and E.~Duchesnay, ``Scikit-learn:
  Machine learning in {P}ython,'' \emph{Journal of Machine Learning Research},
  vol.~12, pp. 2825--2830, 2011.

\end{thebibliography}

\clearpage

\section*{Supplementary Material}

In this supplementary material, we explain how we modify the NPO framework in \cite{SunTayHe2017towards} to include both information privacy and local differential privacy metrics. We call this approach EPIC in \cref{sec:nonparametric} of the main paper. This is a simple extension of the NPO framework and is presented here for completeness. We also include a simulation study to compare the performance of EPIC with empirical optimization frameworks without either the information privacy or local differential privacy constraint.  

\subsection{Empirical Information Privacy and Local Differential Privacy Optimization}\label[appendix]{ap:EPIC}

Following \cite{SunTayHe2017towards}, let $\phi$ be a loss function, $\scH$ be a reproducing kernel Hilbert space with kernel $\kappa(\cdot,\cdot)$, kernel inner product $\ip{\cdot}{\cdot}$, and associated norm $\norm{\cdot}$. We restrict the rule used by the fusion center to infer $H$ and $G$ based on $Z=\bz$ to be of the form $\ip{w}{\Phi(\bz)}$, where $\Phi(\bz)=\kappa(\cdot,\bz)$ is the feature map. We seek to minimize the empirical $\phi$-risk of deciding $H$ while preserving information privacy. 

We consider the following optimization problem:
\begin{subequations}\label{EPIC}
\begin{align}
\min_{w\in\scH,p_{Z\mid X}\in\calQ}&\ F(w,p_{Z\mid X}),\label{objective}\\
\text{s.t.}&\ \min_{v\in\scH}\hat{R}_{\bg}(v,p_{Z\mid X}) \geq \theta,\ \forall \bg\in\calG\backslash\{\bzero\},\label{maryconstraints}\\
&\frac{p_t(z|x)}{p_t(z|x')}\leq e^{\epsilon_{LD}},\nonumber\\
&\hspace{0.8cm} \text{$\forall z\in\calZ$, $x,x'\in\calX$, $t=1,\ldots,s$}, \label{EPIC_LD}
\end{align}
\end{subequations}
where
\begin{align*}
&F(w,p_{Z\mid X})=\ofrac{n}\sum_{i=1}^n\phi\left(h_{i}\ip{w}{\Phi_Q(\bx^i)}\right)+\frac{\lambda}{2}\norm{w}^{2},\\
&\hat{R}_{\bg}(v,p_{Z\mid X})=\ofrac{2}\sum_{\bg'\in\{\bzero,\bg\}}\sum_{i\in\calS_{\bg'}}\frac{\phi(g^{\prime,i}\ip{v}{\Phi_Q(\bx^i)})}{|\calS_{\bg'}|}+\frac{\lambda}{2}\norm{w}^2,\\
&\Phi_Q(\bx)=\sum_{\bz\in\calZ^s}p_{Z\mid X}(\bz|\bx)\Phi(\bz),
\end{align*}
$\lambda >0$, $\theta >0$ is called the information privacy threshold, 
\begin{align*}
g^{\prime,i} =
\begin{cases}
-1, & \text{if } \bg^i = \bzero,\\
1, & \text{otherwise},
\end{cases}
\end{align*}
and 
\begin{align*}
\calS_{\bg'}=\left\{i\in\{1,\ldots,n\}: \bg^i = \bg'\right\}.
\end{align*}
Note that $F(\cdot,\cdot)$ is the empirical $\phi$-risk of detecting $H$ while $\hat{R}_{\bg}(\cdot,\cdot)$ is the empirical (normalized) $\phi$-risk of distinguishing between $G=\bzero$ and $G=\bg$. For convenience, we call \eqref{EPIC} the Empirical information and local differential PrIvaCy (EPIC) optimization.

For a detailed explanation of how the above optimization framework is derived, we refer the reader to \cite{SunTayHe2017towards}. Briefly, we seek to find $p_{Z\mid X}$ such that the empirical risk for detecting $G$ under \emph{any decision rule} adopted by the fusion center is above the information privacy threshold $\theta$. The mapping $p_{Z\mid X}$ is also required to satisfy $\epsilon_{LD}$-local differential privacy in the constraint \cref{EPIC_LD}.

From \cite[Theorem 2]{SunTayHe2017towards}, for each $\epsilon_{LD}$, by choosing $\theta$ appropriately, we can achieve $\epsilon_I$-information privacy for any $\epsilon_I >0$ under mild technical assumptions. However, this trades off the detection error rate for $H$. Therefore, we adopt the same two-step procedure in \cite{SunTayHe2017towards}: 
\begin{enumerate}[(i)]
	\item Determine the largest information privacy threshold $\theta^*$ achievable under additional constraints on $p_{Z\mid X}$ to ensure that the error rate of inferring $H$ remains reasonable. This is achieved through an iterative block Gauss-Seidel method. 
	\item Set a $r\in (0,1)$, which we call the \emph{information privacy threshold ratio}, set $\theta = r\theta^*$ in \eqref{maryconstraints} and use an iterative block Gauss-Seidel method to solve \eqref{EPIC}. 
\end{enumerate}
For the details of this two-step procedure, we again refer the reader to \cite{SunTayHe2017towards}. The only difference with the procedure in \cite{SunTayHe2017towards} is that now we have the additional linear inequality constraints \eqref{EPIC_LD}, which can be easily handled since each step in the block Gauss-Seidel method remains as a convex optimization problem.

\subsection{Simulation Results}\label{EPIC:simulation}

In this subsection, we consider the nonparametric case where the underlying sensor distributions are unknown. We perform simulations to provide insights into the performance of our proposed EPIC approach in \cref{EPIC}.

For simplicity, we use the count kernel in our simulations, which can be computed with a time complexity of $\calO(s|\calY|)$. We choose the logistic loss function as the loss function $\phi$ in our simulations.

Consider a network of $4$ sensors and a fusion center. Each sensor observation $x_t^i$ is generated according to Table~\ref{tab:discreteX}, where $n^i_t$ is uniformly distributed over $\{-2,-1,0,+1,+2\}$. The sensor observation space is $\calX=\{-5,-4,\ldots,5\}$, and the local decision space is chosen to be $\calZ=\{1,2\}$. Conditioned on $(H,G)$, sensor observations are independent of each other. We generate $40$ i.i.d.\ training samples, and apply our proposed approach on the training data to learn the privacy mapping $p_{Z\mid X}$. 

\begin{table}[!ht]
	\centering
	\caption{Sensor observation for different realizations of $(H,G)$.}
		\begin{tabular}{ | c | c | c| c | c | }\hline
    		$(h^i,g^i)$  &$(0,0)$ &$(0,1)$ &$(1,0)$		&$(1,1)$			\\ \hline 									
             $x_t^i$   & $-3+n^i_t$ & $-1+n^i_t$ & $1+n^i_t$ 		& $3+n^i_t$ 	\\ \hline
		\end{tabular}
\label{tab:discreteX}
\end{table}

\begin{figure}[!ht]
\centering
\includegraphics[width=0.9\linewidth]{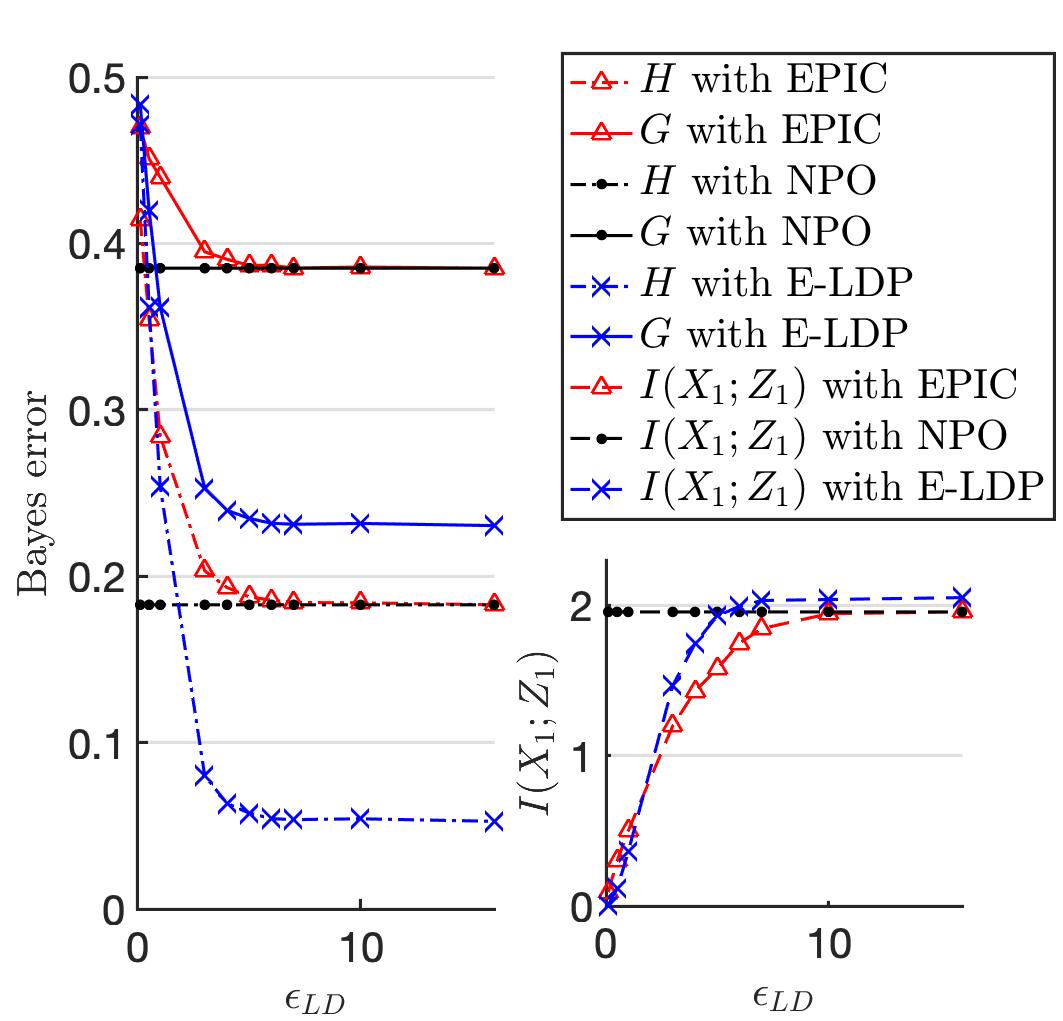}
\caption{Bayes error for detecting $H$ and $G$, and mutual information between $X_1$ and $Z_1$ with different local differential privacy budget $\epsilon_{LD}$. }
\label{fig:dp}
\end{figure}

Fig.~\ref{fig:dp} demonstrates how $\epsilon_{LD}$, the local differential privacy budget, affects the inference privacy, data privacy and utility of these methods. In the simulation, we fix the information privacy threshold ratio $r=0.999$ when setting $\theta=r\theta^*$ in \eqref{maryconstraints}, and the correlation coefficient between $H$ and $G$ is $0.2$. We observe that when $\epsilon_{LD}$ is small, the performance of EPIC is close to the performance of E-LDP, where the Bayes error rates of both hypotheses are close to $0.5$. This is in line with Theorem~\ref{thm:privacy_metrics}\ref{it:ld2i}: a small local differential privacy budget implies information privacy for both hypotheses. With the increase of $\epsilon_{LD}$, the performance of EPIC approaches the performance of NPO, where the error rate of $H$ is low, while that for $G$ is high. However, with E-LDP, the error rate of $G$ also decreases with increasing $\epsilon_{LD}$, which leads to inference privacy leakage. When analyzing the data privacy leakage, we find that $I(X_1;Z_1)$ stays high with NPO, whereas EPIC achieves a reasonable $I(X_1;Z_1)$ by choosing $\epsilon_{LD}$ to be around 5.

\begin{figure}[!ht]
\centering
\includegraphics[width=0.9\linewidth]{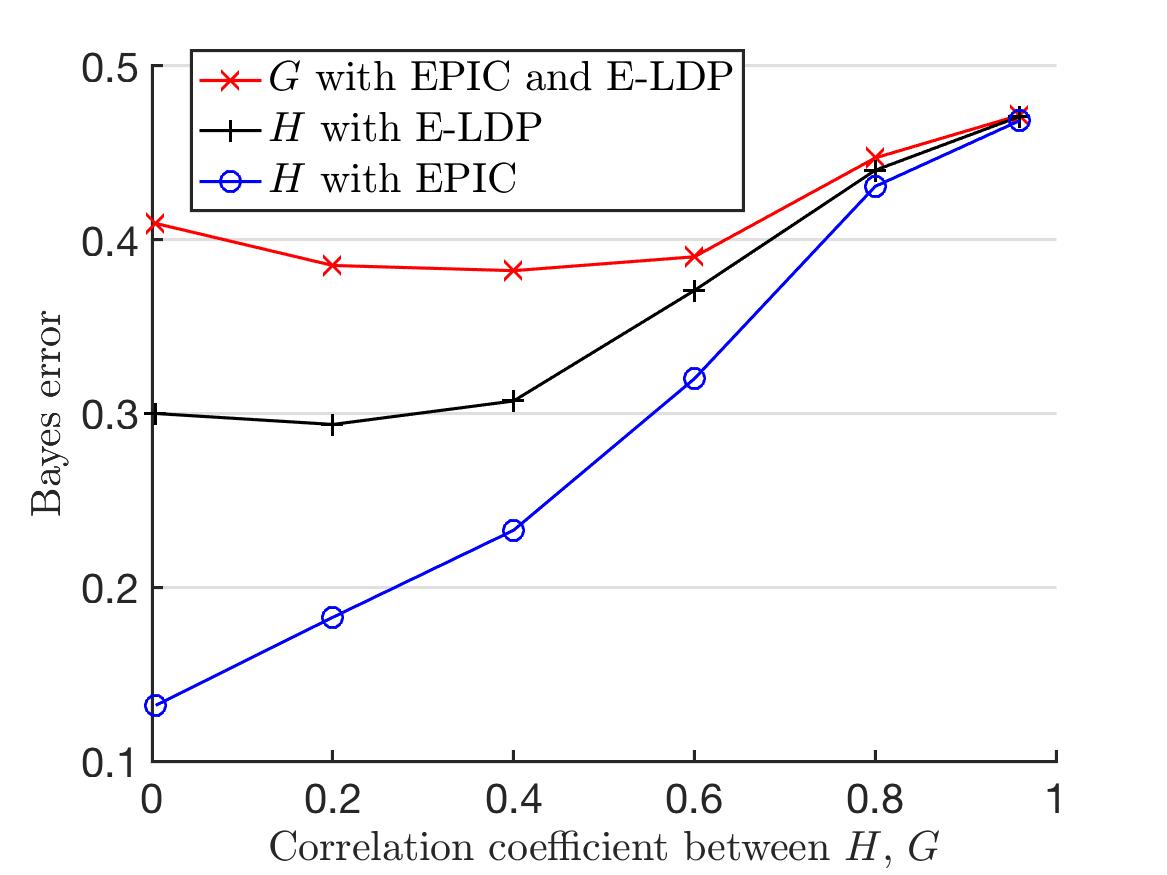}
\caption{Bayes error probability of detecting $H$ and $G$ with varying correlation coefficient between $H$ and $G$. }
\label{fig:correlation}
\end{figure}

Fig.~\ref{fig:correlation} shows how the correlation between $H$ and $G$ affects their Bayes error detection rate. For EPIC, we set $\epsilon_{LD}=5$, and for E-LDP, we find a local differential privacy budget for each correlation coefficient tested that achieves the same error rate for $G$ as in EPIC. We observe that for the same correlation coefficient, the error rate for $H$ is higher in E-LDP compared to that in EPIC. This demonstrates our claim that local differential privacy should not be used to imply information privacy, as it can severely impact the detection error rate for $H$ as well.

\end{document}